%% file: main.tex
\documentclass{article}
\usepackage{iclr2026_conference,times}
\iclrfinalcopy
\usepackage[T1]{fontenc}
\usepackage[utf8]{inputenc}
\usepackage{microtype}
\usepackage{amsmath,amssymb,amsthm,mathtools}
\usepackage{bm}
\usepackage{booktabs}
\usepackage{multirow}
\usepackage{tabularx}
\usepackage{graphicx}
\usepackage{subcaption}
\usepackage{xcolor}
\usepackage{enumitem}
\usepackage{listings}
\usepackage{float}
\usepackage{url}
\usepackage[colorlinks=true,linkcolor={blue!60!black},citecolor={green!45!black},urlcolor={blue!60!black}]{hyperref}
\usepackage[capitalize,nameinlink]{cleveref}
\usepackage{caption}
\setlist{nosep,leftmargin=*}
\graphicspath{{figures/}}
\newcommand{\hop}{\mathrm{hop}}
\newcommand{\G}{\textsf{G}}
\newcommand{\E}{\textsf{E}}
\newcommand{\T}{\textsf{T}}
\newcommand{\ET}{\textsf{ET}}
\newcommand{\grep}{\texttt{grep}}
\newcommand{\readfile}{\texttt{read\_file}}
\newcommand{\listdir}{\texttt{list\_dir}}
\newcommand{\semsearch}{\texttt{semantic\_search}}
\newcommand{\findentity}{\texttt{find\_entity}}
\newcommand{\graphnb}{\texttt{graph\_neighbors}}
\newcommand{\submit}{\texttt{submit}}
\newcommand{\facc}{function Acc@10}
\newcommand{\ci}[2]{[#1,\,#2]}
\newcommand{\figref}[1]{Fig.~\ref{#1}}
\newcommand{\ie}{i.e.}

\DeclareMathOperator{\AUC}{AUC}
\DeclareMathOperator{\logit}{logit}
\theoremstyle{definition}
\newtheorem{definition}{Definition}
\newcommand{\run}[1]{\textsc{#1}}

\title{How Do Small Coding Agents Use Code Graphs and Embeddings?\\[0.15em]
\parbox{\textwidth}{\normalsize\normalfont A depth-stratified study of issue localization on SWE-bench Verified, with tool ablations, a 0.6B--32B scale curve, downstream repair, and linear probes of the repository graph}}

\author{Saurav Bellam \\
The University of Texas at Austin \\
\texttt{sauravbellam@utexas.edu}
}

\begin{document}
\maketitle

\begin{abstract}
\input{sections/abstract}
\end{abstract}

\input{sections/intro}
\input{sections/related}
\input{sections/depth}
\input{sections/agent}
\input{sections/results_tools}
\input{sections/probes}
\input{sections/discussion}
\input{sections/conclusion}

\section*{Reproducibility and data release}
Everything behind this paper is public. The depth-labelled split (labels for all 500 tasks under every variant, a dataset card, the labelling scripts and the \run{DepthGEE}/\run{LabelVar} refit output) is at \url{https://github.com/SauravBellam/swebench-verified-hop-depth} (CC~BY~4.0) and is also included as ancillary files of this arXiv submission (\texttt{anc/}). The complete package---the agent and probe source code, every raw result file (episode logs, conversation traces, per-pair probe scores and SWE-bench harness reports; about 60\,MB for the primary runs and 140\,MB for the \run{ScaleCurve} runs under \texttt{results/scale\_curve/}, with the exact notebooks, environment records and serving commands that produced them), and the analysis scripts that regenerate every table and figure from them (\texttt{analysis/scale\_curve.py} for \cref{sec:scalecurve})---is at \url{https://github.com/SauravBellam/small-agents-code-graphs} (Apache~2.0).

\bibliographystyle{iclr2026_conference}
\bibliography{refs}

\clearpage
\appendix
\small\linespread{0.93}\selectfont\setlength{\intextsep}{5pt}\setlength{\textfloatsep}{8pt}
\input{sections/appendix}

\end{document}

%% file: sections/abstract.tex
Coding agents are given embedding search and code-graph tools on the premise that a repository's structure is something the model lacks and the tool supplies. We test that premise on SWE-bench Verified with small open-weight models. First, we label all 500 tasks with a deterministic hop depth: the shortest code-graph path from code the issue mentions to code the gold patch changes (186/128/78/76 tasks at hop 0/1/2/3+). With repository, tool-set and patch-size controls, each extra hop multiplies the odds of correct function-level localization by 0.47--0.55 for Qwen3--4B, Qwen3--8B, Gemma~4 E4B and Qwen3--32B, and by 0.50--0.67 at each of six Qwen3 sizes from 0.6B to 32B. Second, we ablate a localization agent's tool set (grep only; +embedding search; +graph tools; all) on a depth-balanced 160-task split and on all 500 tasks for ten models (79k episodes). Structured tools double grep-only accuracy for Qwen3--8B and Gemma~4 E4B; their gain peaks at 17--21 points at 1.7--4B and falls to 4--5 points at 32B (odds ratio 0.77--0.82 per doubling of parameters). From 8B up, about half of that gain is the form of the submission rather than finding the target; the fall from 14B to 32B is a fall in tool uptake at equal grep-only accuracy; and the same 4B size gains or does not gain from a tool depending on the checkpoint. Traces show the graph tools' gain is name lookup at hop 0 and one expansion at hop 1: agents chain expansions in 12--14\% of episodes and never flatten the depth gradient. A fixed repair model resolves 6--13 of 160 tasks from these localizations (oracle 26). Third, linear probes find direct call edges decodable under a within-anchor, degree-stratified protocol (AUC 0.74), but below a lexical baseline (0.99), with two-hop edges at 0.57. Small agents carry a one-hop, largely lexical picture of their repository, in weights and in use. We release the labels, scripts, logs and traces.

%% file: sections/intro.tex
\section{Introduction}
\label{sec:intro}

Resolving a GitHub issue in a large repository starts with finding the code that has to change. Agentic systems for this \emph{localization} step increasingly ship with structured tools: dense retrievers over function bodies, and query interfaces over a static code graph whose nodes are files, classes and functions and whose edges record containment, imports, calls and inheritance \citep{locagent,orcaloca,cosil,kgcompass}. The working assumption behind these tools is simple. An issue names some code; the fix lives some distance away along the dependency structure; a plain text search cannot bridge that distance, so the agent needs either a semantic index (to jump directly to code that \emph{means} the right thing) or a graph (to \emph{walk} to it). LocAgent \citep{locagent} made the assumption measurable by scoring SWE-bench tasks by the hop distance between the functions an issue mentions and the functions the patch edits, and reported that graph-guided agents degrade least as that distance grows. CodeAnchor \citep{codeanchor} observed that SWE-bench Lite and Verified contain mostly one- and two-hop faults, and argued that evaluating structure-aware agents properly ``requires a benchmark that sorts tasks by structural depth''.

This paper takes that argument seriously and asks three questions that the existing literature leaves open, using small open-weight models where the behaviour is cheap to measure many times over and where tool support plausibly matters most.

\paragraph{Part 1. Is structural depth a usable difficulty axis?} We compute, for every SWE-bench Verified task, a deterministic hop depth on a LocAgent-style graph built at the task's base commit (\cref{sec:depth}). Unlike LocAgent's measurement, which extracted mentions with GPT-4o, ours uses only regular expressions and the Python \texttt{ast} module, so it can be re-run, audited and varied. We ask how the 500 tasks are distributed over depth, whether depth predicts whether an agent localizes the fix after controlling for repository, tool set, number of gold functions and patch size (\run{DepthGEE}), how sensitive the label is to the two measurement choices that matter---how calls are resolved and how mentions are extracted (\run{LabelVar})---and we release the labelled split.

\paragraph{Part 2. What do \grep, embedding search and graph tools each contribute?} On a 160-task subset balanced over depth we run a localization-only agent whose arms differ \emph{only} in their tool set: \G\ (\grep, \readfile, \listdir), \E\ (\G\ + embedding search over function and class texts), \T\ (\G\ + a name lookup and a one- or two-hop neighbourhood query over the graph) and \ET\ (all tools) (\cref{sec:agent}). We measure accuracy by depth, run-to-run variance under sampling, which tool first exposes the gold code and whether exposure converts into a correct submission, sensitivity to retrieval quality, the effect of forcing the structured tool to be used, the effect of a larger step and context budget, and---because localization is only useful if it leads to a fix---how many issues one fixed repair model resolves when given each arm's localizations (\cref{sec:tools}). We then rerun the four arms on all 500 tasks across a model-size curve: six Qwen3 sizes from 0.6B to 32B with five decodings each, an FP8 precision control, and three Gemma~4 sizes (\cref{sec:scalecurve}).

\paragraph{Part 3. Does the model represent the repository's graph internally?} The ``Thousand-Graph Hypothesis'' \citep{thousandgraph} proposes that an entity-only index suffices because self-attention materialises whatever relation a task needs; a recent probe shows that the \emph{runtime tool-call} dependency structure of an agent is linearly decodable from Qwen3-32B residual streams \citep{toolcallprobe}. Neither probes the repository's own call and inheritance graph. We do, with linear probes on definition-level activations, first in static file contexts (\run{GlobalProbe}, \run{AnchorProbe}) and then inside real agent conversations (\run{ConvProbe}). The design is driven by a negative result we obtained early on: a global pair probe reaches AUC 0.75--0.80 for call edges, but so does a probe that sees only one of the two definitions, and so does the in-file in-degree of the callee. Linked and unlinked pairs differ in properties of a \emph{single} node. We therefore introduce a within-anchor ranking protocol with an in-degree-binned stratification that a node-level shortcut cannot pass, with bootstrap intervals over anchors and a lexical name-match baseline (\cref{sec:probes}).

\paragraph{Findings.} The answers are consistent across two model families and six sizes, and point in one direction.
\begin{itemize}
\item Hop depth is a strong, controlled predictor of failure. Each extra hop multiplies the odds of correct function-level localization by 0.47 (Qwen3-4B, three runs), 0.47 (Qwen3-8B), 0.54 (Gemma~4 E4B, three runs) and 0.55 (Qwen3-32B, all 500 tasks), with all $p\le10^{-3}$; across the scale curve the per-hop odds ratio is 0.50--0.67 at every size and barely changes with size. The label barely moves under alternative call resolution ($\kappa=0.96$--$0.98$). It is, however, sensitive to mention extraction: a strict extractor that counts only back-ticked code and traceback frames leaves 311 of 500 tasks without a mention.
\item For Qwen3-8B and Gemma~4 E4B, \grep\ alone is a weak baseline (12.8\% and 9.6\% \facc) that any structured tool roughly doubles. For Qwen3-4B the embedding tool adds nothing ($-1.1$ points) and the graph tools add $+8.1$ points ($\ci{3.0}{13.5}$). In every model, the graph-tool advantage is largest at hop 0, where the issue names the target and \findentity\ resolves the name directly, and at hop 1, where one \graphnb\ expansion reaches it; at hop 2+ the gain shrinks or vanishes, and in a GEE interaction model the depth slope of \T\ is never significantly flatter than that of \G\ (it is steeper in the Qwen models) while \E\ has the flattest slope. Small graph-tool agents do not walk the graph; they look names up and expand once. Over the scale curve, the structured tools' gain over \grep\ rises from 7--9 points at 0.6B to 17--21 at 1.7--4B and falls to 4--5 at 32B (odds ratio 0.77--0.82 per doubling of parameters); from 8B up about half of that gain is the \emph{form} of the submission (exact ids) rather than discovery; the fall from 14B to 32B happens at equal \grep-only accuracy and is a fall in tool uptake (\findentity\ called in 96\% of episodes at 14B, 45\% at 32B), not an FP8 artifact; and the same 4B size gains $+17.6$ points from embedding search under one checkpoint and nothing under another.
\item Grep-only Qwen3-8B collapses into search loops (10 \grep\ calls per episode, 40\% of episodes hit the step cap, 41\% never submit); a 50-step, 40k-token budget does not rescue it. Graph tools also change the \emph{form} of what the agent submits: 85\% of its top-5 function strings are exact entity ids, against 47\% with \grep\ only, which matters downstream.
\item Downstream, the localizations of \G/\E/\T/\ET\ lead a fixed Qwen3-8B repair prompt to 6/13/12/13 resolved tasks of 160 (oracle locations: 26). Resolution is almost entirely conditional on correct localization (24--27\% versus 1--2\%).
\item Inside the model, direct call edges are linearly decodable within anchor and in-degree bin (AUC 0.74 and 0.68 on the two held-out folds for Qwen3-4B; 0.73 and 0.66 for Qwen3-8B) and remain so inside agent conversations (0.61--0.66) including for edges no tool printed (0.59--0.60); but a one-line lexical baseline reaches 0.99, two-hop edges with the lexical cue removed reach only 0.57, and inheritance is not decodable beyond one-sided information. What is in the weights is a shallow, mostly lexical picture of direct dependencies.
\end{itemize}

\paragraph{Contributions.} (i) A deterministic, released hop-depth labelling of SWE-bench Verified with a validated difficulty analysis and a robustness study of the label itself (\cref{sec:depth}). (ii) A controlled, depth-stratified tool ablation with trace-level mechanism analysis, variance, forced-use, budget, retrieval-quality and downstream-repair experiments, and a full-benchmark scale curve over ten models (six Qwen3 sizes from 0.6B to 32B with five decodings each, an FP8 control and three Gemma~4 sizes; 79k episodes) with a canonical rescoring that separates discovery from the form of the submission (\cref{sec:tools}). (iii) A within-anchor, degree-stratified probing protocol for relational structure, with the negative control that motivates it, applied both to static code and to activations taken inside agent conversations (\cref{sec:probes}). (iv) A synthesis: with small models, the benefit of ``structure'' tools is exact-name resolution plus one hop, which is also the extent of the structure the models appear to carry internally, and which shrinks with scale as the models stop calling the tools; deeper tasks are failed by every arm at every size (\cref{sec:discussion}).

All numbers in this paper are computed from the episode logs, traces, probe scores and harness reports described in the reproducibility statement; the depth labels, source code, raw result files and analysis scripts are public (see the reproducibility statement).

%% file: sections/related.tex
\section{Related work}
\label{sec:related}

\paragraph{Graph tools versus one-shot retrieval as depth grows.}
LocAgent \citep{locagent} builds a heterogeneous code graph (file, class, function nodes; contain, import, invoke, inherit edges) and gives an agent tools to search entities, traverse the graph and retrieve code. Its difficulty analysis (App.\ C.1.2) measures the shortest hop between the functions an issue names and the patched functions, finds accuracy falling with depth for every method, and reports that embedding retrievers (E5, CodeRankEmbed) and Agentless \citep{agentless} fall furthest at hop 3+ while graph-tool agents degrade least. SWE-Explore \citep{sweexplore} finds that sparse retrievers and a lightweight static-embedding retriever ``remain close to Random on most metrics'' whereas all agentic explorers score much higher. A family of graph-guided localizers follows the same design: OrcaLoca \citep{orcaloca}, iterative code-graph search \citep{cosil}, repository-aware knowledge graphs \citep{kgcompass}, conceptual-knowledge extraction \citep{repolens}, unified knowledge graphs for multilingual issue resolution \citep{prometheus}, tool-integrated RL for repository search \citep{tooltrain}, repository memory \citep{repomemory}, multi-language localization \citep{multicolor}, location inference for task-level understanding \citep{deepdiscovery} and exploration structure for multi-file changes \citep{explorationstructure}. Our study differs in three ways: the models are small (4--8B), the tool sets are ablated one at a time on the same tasks with the same harness, and the depth axis is used as a stratification variable with controls rather than as a post-hoc breakdown. Our result on depth differs from LocAgent's at hop 3+: with small models the graph arm's depth slope is no flatter than \grep's.

\paragraph{Embeddings that know about structure.}
SpIDER \citep{spider} adds graph-neighbourhood information to dense retrieval for issue localization; Meta-RAG \citep{metarag} retrieves over LLM summaries of code; ProjAgent \citep{projagent} retrieves by procedural similarity for repository-level generation; CORE-Bench \citep{corebench} and Agent Retrieval Bench \citep{agentretrievalbench} evaluate repository retrieval for agents; \citet{deepagenticsearch} compare a vector index with grep-style agentic search for repository QA. Our embedding arm uses a standard instruction-tuned embedder (Qwen3-Embedding-0.6B) over function and class texts without structural augmentation, which is the configuration whose contribution we want to isolate; the \run{WeakEmb} experiment (\cref{sec:b6}) swaps in a weak static embedder to measure how much of the arm's effect is retrieval quality.

\paragraph{How structure is shown to the agent.}
RepoNav \citep{reponav} reorganises flat snippet retrieval into a file-centred scaffold and finds that the \emph{organisation} of structure helps beyond its volume (File-Only 32.6\% $\to$ Inline 47.0\% $\to$ Tree 52.1\% function Acc@5). CodeAnchor \citep{codeanchor} and LARGER \citep{larger} inject structure into \grep\ results instead of adding tools. Rendering the repository as an image also helps \citep{seerepo}; incremental code indexes can silently serve wrong graphs \citep{codenib}; context files did not measurably move correctness in a two-agent ablation \citep{contextfiles}; ontology-grounded project memory exposes design rationale as a knowledge graph \citep{moosedev}. DeepRepoQA \citep{deeprepoqa} and agentic code reasoning with semi-formal certificates \citep{agenticcodereasoning} target understanding rather than localization. Our \T\ arm presents structure in the plainest way (one line per edge, truncated at 50 lines), and our trace analysis (\cref{sec:b5}) shows which of those lines the agent actually acts on.

\paragraph{Do models represent the graph internally?}
The Thousand-Graph Hypothesis \citep{thousandgraph} proposes that an entity-only index suffices because self-attention materialises task-specific relations; its two-layer index reaches 95.6\% on SWE-bench Verified with DeepSeek-V4-Flash, but the author notes this ``does not directly verify the latent graph realized inside the model''. \citet{toolcallprobe} decode an agent's \emph{runtime tool-call} dependency graph from Qwen3-32B residual streams above random-label and positional controls, and are careful that their ``claims concern representation, not behavioural control''. Neither probes the repository's call or import graph. Our probes do, and the protocol we arrive at---within-anchor ranking with degree stratification, a permuted-label control and a lexical baseline---is a direct response to the node-level confounds we document in \cref{sec:a1}. The general concern that pair probes can be solved by single-item features is the relational analogue of the selectivity argument of \citet{hewitt2019control}.

\paragraph{Controlled difficulty.}
CABRA \citep{cabra} builds synthetic tasks as call-graph transformations and finds that agents stay near-perfect as task size grows because they offload to tools such as \grep, and that tool-call counts predict SWE-bench Verified accuracy better than lines edited. CodeAnchor \citep{codeanchor} observes that SWE-bench Lite and Verified mostly contain one- and two-hop faults. Our depth labels quantify that observation on Verified (67\% of tasks with a mention are at hop 0 or 1) and supply the sorted benchmark CodeAnchor asks for. Graph retrieval for repository-level code \emph{generation} is adjacent but not our topic \citep{graphcodeagent,grace,partialdepgraph,crossrepo,featlens,kgasfiles}.

%% file: sections/depth.tex
\section{Part 1: Hop depth as a difficulty axis}
\label{sec:depth}

\subsection{Code graph}
\label{sec:graph}

For each SWE-bench Verified \citep{swebench,swebenchverified} instance $i$ we check out the repository at its \texttt{base\_commit} and parse every \texttt{.py} file with the Python \texttt{ast} module (files that fail to parse contribute a file node only). The result is an undirected, typed multigraph
\begin{equation}
\begin{aligned}
\mathcal{G}_i&=(V_i,E_i),\qquad V_i = V^{\mathrm{file}}\cup V^{\mathrm{class}}\cup V^{\mathrm{func}},\\
E_i&\subseteq \tbinom{V_i}{2}\times\{\textsc{contain},\textsc{import},\textsc{invoke},\textsc{inherit}\},
\end{aligned}
\end{equation}
following LocAgent \citep{locagent}. A function or class node is identified by its path and qualified name, \texttt{path/to/mod.py:Class.method}; its span $[s_v,e_v]$ is the line range of the definition including decorators. Edges are produced as follows.
\begin{itemize}
\item \textsc{contain}: file $\to$ top-level definition, class $\to$ member, for definitions nested in \texttt{if}/\texttt{try}/\texttt{with}/\texttt{for}/\texttt{while} blocks as well.
\item \textsc{import}: file $\to$ file for every \texttt{import m} or \texttt{from m import x}, resolving relative imports from the importing module's package; a dotted module is shortened to its longest prefix that is a file in the snapshot, and self-imports are dropped.
\item \textsc{invoke}: function $\to$ definition for every \texttt{ast.Call} whose callee is a bare name or an attribute. Resolution is by name with a fixed preference order: a \texttt{self.x()}/\texttt{cls.x()} call prefers methods of the enclosing class; otherwise definitions in the same file; otherwise definitions in the module that \texttt{from}-imported the name; otherwise every definition of that name in the repository, provided there are at most $K_{\mathrm{res}}=5$ of them (ambiguous names are dropped).
\item \textsc{inherit}: class $\to$ class for each base expressed as a name or attribute, resolved the same way and kept only if the target is a class.
\end{itemize}
Snapshots are large: a mean of 26{,}242 nodes and 232{,}915 edges (median 34{,}919 and 183{,}269; maximum 45{,}677 and 1{,}024{,}494). Name-based resolution is deliberately cheap; its consequences are measured in \cref{sec:c3}.

\subsection{Targets, mentions and depth}
\label{sec:depthdef}

\begin{definition}[Targets]
Let $P_i$ be the gold patch. For each removed line at old-side line number $\ell$ in file $f$, and for each added line at the old-side position before it, the target is the innermost definition $v$ with $f_v=f$ and $s_v\le \ell\le e_v$ (the file node if the line is at module level). $T_i\subseteq V_i$ is the set of all such targets.
\end{definition}

\begin{definition}[Mentions]
Let $D_i$ be the issue text. A mention is a token produced by one of six regular-expression patterns: back-ticked spans, identifiers immediately followed by \texttt{(}, traceback frames \texttt{File "...py", line n, in f}, dotted names, \texttt{snake\_case} names and \texttt{CamelCase} names; every token is further split into identifier-like pieces. A token ending in \texttt{.py} maps to every file whose path equals or ends with it; any other token of length $\ge 3$ maps to every definition with that short name, unless more than $K_{\mathrm{men}}=20$ definitions share it, in which case it is treated as too generic and discarded. $M_i=\{m_1,\dots,m_k\}$ is the set of mentions, each a set of nodes $m_j\subseteq V_i$.
\end{definition}

\begin{definition}[Hop depth]
Let $d_{\mathcal{G}}(u,v)$ be the undirected shortest-path length in $\mathcal{G}_i$ over a chosen edge-type subset. The depth of a task is
\begin{equation}
\hop_i=\min_{m_j\in M_i}\ \min_{u\in m_j,\;v\in T_i} d_{\mathcal{G}_i}(u,v),
\label{eq:hop}
\end{equation}
computed by a breadth-first search from $T_i$. The bucket is $\mathrm{hop}0/\mathrm{hop}1/\mathrm{hop}2/\mathrm{hop}3{+}$ for $\hop_i=0,1,2,\ge3$; \emph{no\_mention} if $M_i=\varnothing$ or $T_i=\varnothing$; \emph{unreachable} if no mention is connected to any target. For comparison we also compute LocAgent's $\delta_i=\frac{1}{|M_i|}\sum_{j}\frac{1}{1+d_j}$ with $d_j$ the per-mention minimum distance, and the mean per-mention distance.
\end{definition}

Hop 0 means the issue names a definition the patch changes (every hop-0 task has at least one mentioned target, and no task in a deeper bucket does). Hop 1 means the nearest mentioned definition is one containment, import, call or inheritance edge away: a method of a mentioned class, a function in a mentioned file, a callee of a mentioned function. Depth is computed from the issue text and the gold patch, so it must never be shown to a system under evaluation; it is a property of the task, not of any agent's path.

\subsection{Distribution of depth on SWE-bench Verified}
\label{sec:c1}

\begin{figure}[t]
\centering
\includegraphics[width=\linewidth]{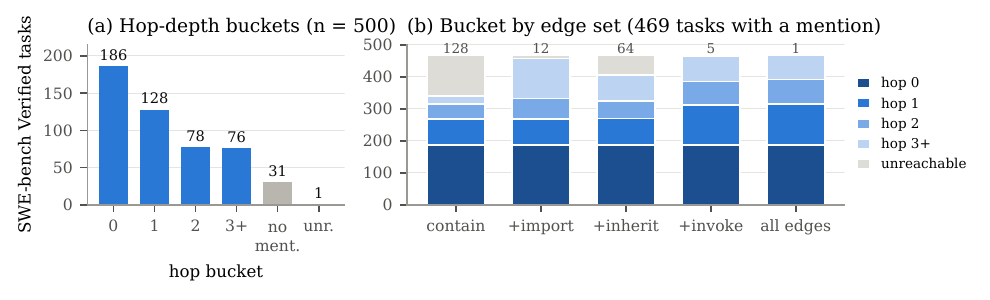}
\caption{\textbf{Hop depth on SWE-bench Verified.} (a) Default label (all four edge types). (b) The same 469 tasks with a mention and a target, re-labelled with containment edges only and with containment plus one other edge type; grey segments are tasks whose mentions are disconnected from the targets under that edge set. Calls are the edge type that connects most tasks; imports and inheritance alone leave 12 and 64 tasks unreachable respectively.}
\label{fig:hopdepth}
\end{figure}

\figref{fig:hopdepth}a gives the distribution: 186 tasks at hop 0 (37.2\%), 128 at hop 1 (25.6\%), 78 at hop 2 (15.6\%), 76 at hop 3+ (15.2\%), 31 without a usable mention (6.2\%) and one unreachable. Two thirds of the tasks with a mention (314/469) are therefore within one edge of something the issue names, which quantifies CodeAnchor's observation that the benchmark ``mostly contains 1--2 hop faults'' \citep{codeanchor}. The distribution differs by repository (\cref{tab:byrepo}, appendix): Django, which contributes 231 tasks, is 87/64/29/38 over the four buckets, while Sphinx is skewed deep (2/5/11/17) and scikit-learn shallow (20/9/3/0).

\figref{fig:hopdepth}b shows what each edge type contributes. With \textsc{contain} edges only, 128 of the 469 tasks are unreachable (the mention and the target are in different files); adding \textsc{invoke} edges connects all but 5, adding \textsc{import} all but 12, adding \textsc{inherit} all but 64; all four together leave one. The mean depth under all edges is 1.17 (median 1); under containment plus calls 1.18; plus imports 1.57; plus inheritance 1.64. Along the default label, the mean number of targets falls with depth (2.34, 2.44, 1.73, 1.57 at hop 0--3+) and so does the number of mentions (8.56, 6.48, 7.15, 3.67): deep tasks are typically single-function fixes described by issues that name little code. LocAgent's $\delta$ averages 0.49, 0.35, 0.27 and 0.21 over the buckets.

\subsection{Does depth predict localization failure? (\run{DepthGEE})}
\label{sec:c2}

We fit a population-averaged logistic model on the episodes of the agent experiments of \cref{sec:tools}. For episode $e$ of task $i$ with tool set $a_e$, outcome $y_e\in\{0,1\}$ is \facc\ (\cref{sec:metrics}), and
\begin{equation}
\logit\Pr(y_e=1)=\beta_0+\beta_{\hop}\,\hop_i^{(3)}+\bm{\beta}_{\mathrm{arm}}^{\top}\mathbf{1}[a_e]+\bm{\beta}_{\mathrm{repo}}^{\top}\mathbf{1}[r_i]+\beta_{g}\,n^{\mathrm{gold}}_i+\beta_{p}\log(1+L_i),
\label{eq:gee}
\end{equation}
where $\hop_i^{(3)}=\min(\hop_i,3)$, $\mathbf{1}[a_e]$ is a treatment coding of the arm with \G\ as reference, $\mathbf{1}[r_i]$ codes the repository with repositories contributing fewer than 40 episodes pooled into ``other'', $n^{\mathrm{gold}}_i$ is the number of function-level gold targets and $L_i$ the number of changed lines in the gold patch. The model is estimated by generalised estimating equations \citep{liangzeger1986} as implemented in \texttt{statsmodels} \citep{statsmodels} (binomial family, exchangeable working correlation) with episodes clustered by task, so that repeated runs of the same task and the four arms on the same task do not count as independent. We report $\mathrm{OR}_{\hop}=\exp\beta_{\hop}$, the multiplicative change in the odds of success per additional hop, with a robust 95\% interval, with and without the control block. Tasks a label variant places in \emph{no\_mention} or \emph{unreachable} are dropped from that fit.

\begin{table}[t]
\centering\small
\caption{\textbf{\run{DepthGEE} / \run{LabelVar}: odds ratio per additional hop} for \facc, from the GEE model of \cref{eq:gee}, for the default label and three measurement variants. ``plain'' includes only hop and arm; ``controls'' adds repository, number of gold functions and log patch size. Qwen3-4B and Gemma~4 E4B pool three runs (temperature 0 and two seeds at 0.7) on the 160-task subset; Qwen3-8B uses one temperature-0 run on the subset; Qwen3-32B uses one temperature-0 run on all 500 tasks (457 bucketed tasks with function-level gold); the ``(500)'' rows refit the Qwen3-4B and Qwen3-8B greedy runs on the same 500 tasks (\cref{sec:scale}). $n_{\mathrm{ep}}$ is the number of episodes, $n_{\mathrm{task}}$ the number of tasks.}
\label{tab:c2}
\resizebox{\linewidth}{!}{%
\begin{tabular}{llrrcccc}
\toprule
Model & Label variant & $n_{\mathrm{ep}}$ & $n_{\mathrm{task}}$ & \multicolumn{2}{c}{plain} & \multicolumn{2}{c}{controls}\\
\cmidrule(lr){5-6}\cmidrule(lr){7-8}
 & & & & OR [95\% CI] & $p$ & OR [95\% CI] & $p$\\
\midrule
Qwen3-4B & default & 1872 & 156 & 0.523 \ci{0.401}{0.682} & $1.8{\times}10^{-6}$ & 0.472 \ci{0.345}{0.644} & $2.4{\times}10^{-6}$\\
 & strict\_resolve & 1872 & 156 & 0.537 \ci{0.414}{0.696} & $2.6{\times}10^{-6}$ & 0.474 \ci{0.353}{0.636} & $6.5{\times}10^{-7}$\\
 & loose\_resolve & 1872 & 156 & 0.518 \ci{0.393}{0.682} & $2.8{\times}10^{-6}$ & 0.467 \ci{0.338}{0.645} & $3.7{\times}10^{-6}$\\
 & strict\_mentions & 612 & 51 & 0.657 \ci{0.428}{1.008} & 0.054 & 0.553 \ci{0.344}{0.888} & 0.014\\
\midrule
Qwen3-8B & default & 624 & 156 & 0.547 \ci{0.403}{0.744} & $1.2{\times}10^{-4}$ & 0.468 \ci{0.334}{0.655} & $9.8{\times}10^{-6}$\\
 & strict\_resolve & 624 & 156 & 0.546 \ci{0.408}{0.732} & $4.9{\times}10^{-5}$ & 0.464 \ci{0.338}{0.636} & $1.9{\times}10^{-6}$\\
 & loose\_resolve & 624 & 156 & 0.549 \ci{0.400}{0.753} & $2.0{\times}10^{-4}$ & 0.470 \ci{0.333}{0.663} & $1.7{\times}10^{-5}$\\
 & strict\_mentions & 204 & 51 & 0.565 \ci{0.351}{0.910} & 0.019 & 0.419 \ci{0.218}{0.803} & 0.0088\\
\midrule
Gemma 4 E4B & default & 1872 & 156 & 0.571 \ci{0.430}{0.758} & $1.1{\times}10^{-4}$ & 0.544 \ci{0.387}{0.766} & $4.9{\times}10^{-4}$\\
 & strict\_resolve & 1872 & 156 & 0.558 \ci{0.427}{0.729} & $1.9{\times}10^{-5}$ & 0.529 \ci{0.385}{0.729} & $9.7{\times}10^{-5}$\\
 & loose\_resolve & 1872 & 156 & 0.579 \ci{0.433}{0.774} & $2.3{\times}10^{-4}$ & 0.551 \ci{0.388}{0.783} & $8.8{\times}10^{-4}$\\
 & strict\_mentions & 612 & 51 & 0.767 \ci{0.500}{1.176} & 0.22 & 0.681 \ci{0.444}{1.044} & 0.078\\
\midrule
Qwen3-4B (500) & default & 1828 & 457 & 0.537 \ci{0.459}{0.628} & $8.4{\times}10^{-15}$ & 0.469 \ci{0.393}{0.561} & $1.0{\times}10^{-16}$\\
 & strict\_mentions & 740 & 185 & 0.714 \ci{0.582}{0.875} & 0.0012 & 0.666 \ci{0.538}{0.824} & $1.8{\times}10^{-4}$\\
Qwen3-8B (500) & default & 1828 & 457 & 0.535 \ci{0.448}{0.639} & $4.7{\times}10^{-12}$ & 0.478 \ci{0.396}{0.578} & $2.0{\times}10^{-14}$\\
 & strict\_mentions & 740 & 185 & 0.659 \ci{0.527}{0.825} & $2.7{\times}10^{-4}$ & 0.654 \ci{0.513}{0.833} & $5.9{\times}10^{-4}$\\
\midrule
Qwen3-32B & default & 1828 & 457 & 0.593 \ci{0.500}{0.704} & $2.1{\times}10^{-9}$ & 0.552 \ci{0.453}{0.671} & $2.7{\times}10^{-9}$\\
 & strict\_resolve & 1828 & 457 & 0.587 \ci{0.498}{0.692} & $2.3{\times}10^{-10}$ & 0.541 \ci{0.449}{0.653} & $1.4{\times}10^{-10}$\\
 & loose\_resolve & 1828 & 457 & 0.592 \ci{0.496}{0.706} & $5.1{\times}10^{-9}$ & 0.549 \ci{0.449}{0.673} & $6.6{\times}10^{-9}$\\
 & strict\_mentions & 740 & 185 & 0.784 \ci{0.627}{0.980} & 0.033 & 0.817 \ci{0.632}{1.056} & 0.12\\
\bottomrule
\end{tabular}}

\end{table}

\cref{tab:c2} shows that depth is a strong predictor. Under the default label with controls, each additional hop multiplies the odds of correct function-level localization by 0.472 $\ci{0.345}{0.644}$ for Qwen3-4B, 0.468 $\ci{0.334}{0.655}$ for Qwen3-8B, 0.544 $\ci{0.387}{0.766}$ for Gemma~4 E4B and 0.552 $\ci{0.453}{0.671}$ for Qwen3-32B on all 500 tasks. Adding the controls makes the estimate \emph{stronger}, not weaker, in every model, so depth is not a proxy for repository, patch size or number of targets. The control coefficients themselves are small and not significant ($\mathrm{OR}$ per gold function 0.44--0.72, per log patch line 0.58--1.01, all $p>0.07$ except patch size for the 32B model, 0.67, $p=0.022$), and the arm coefficients reproduce the ablation results of \cref{sec:tools} (for Qwen3-8B: \E\ 2.83, \T\ 2.40, \ET\ 3.43 relative to \G, all $p<10^{-3}$). The appendix (\cref{tab:gee_extra}) reports an interaction model, $\hop\times\mathrm{arm}$, which we use in \cref{sec:b1} to ask whether any tool set flattens the depth gradient. The scale curve of \cref{sec:scalecurve} extends the estimate to six Qwen3 sizes on all 500 tasks: with the arm as a control, the odds ratio per hop is 0.50--0.67 at every size from 0.6B to 32B, all $p<10^{-6}$ (\cref{tab:scale_bucket}).

\subsection{Sensitivity of the label to measurement choices (\run{LabelVar})}
\label{sec:c3}

The two choices most likely to move a label are how a call is resolved and what counts as a mention. We recompute every label under three variants and refit \cref{eq:gee} for each (\cref{tab:c2}, lower rows of each block):
\begin{itemize}
\item \texttt{strict\_resolve}: $K_{\mathrm{res}}=1$; a call or base resolves by repository-wide name only if the name is unique (same-file and same-class resolution still apply).
\item \texttt{loose\_resolve}: $K_{\mathrm{res}}=20$.
\item \texttt{strict\_mentions}: only back-ticked spans and traceback frames count as mentions, and $K_{\mathrm{men}}=5$.
\end{itemize}
\textbf{Call resolution barely matters.} Strict resolution keeps the bucket of 97.4\% of the 500 tasks (Cohen's $\kappa=0.965$; over the 468 bucketed tasks, 97.2\%, $\kappa=0.961$) and moves 34 tasks deeper and none shallower, as it must since it only removes edges; loose resolution keeps 98.6\% ($\kappa=0.981$) and moves 11 tasks shallower. The odds ratios change by at most 0.02 in any model. \textbf{Mention extraction matters.} The strict extractor leaves 311 tasks without any mention: most issues describe their problem with prose identifiers (``the \texttt{Model.save} method'', ``\texttt{QuerySet.values\_list}'') rather than back-ticks or tracebacks. Where it keeps a mention, it agrees with the default bucket on 65.2\% of tasks ($\kappa=0.527$) and is never shallower (71 tasks move deeper, because a dropped mention can only lengthen the minimum path). On the 51 evaluated tasks that keep a mention, the controlled odds ratios are 0.55, 0.42 and 0.68 (the last, for Gemma, not significant at $n=612$ episodes), and 0.82 for Qwen3-32B on the 185 tasks that keep a mention ($p=0.12$). The default extractor is therefore the one we release and use throughout, and the labels should be read as \emph{optimistic}: an ambiguous name counts its closest definition and dynamic dispatch is invisible, so the true distance is at least the labelled one.

\subsection{Release}
The depth-labelled split (\texttt{swebench-verified-hop-depth}\footnote{\url{https://github.com/SauravBellam/swebench-verified-hop-depth}}, also included with this submission as ancillary files) contains one row per task with the default bucket and hop, the hop under each edge set, LocAgent's $\delta$, graph sizes, and the three variant labels, together with the labelling scripts (\texttt{hop\_depth.py}, \texttt{c3\_robustness.py}), the analysis script (\texttt{c2\_depth\_model.py}), the refit output and a dataset card with the caveats above. \cref{sec:app_depth} lists the per-repository distribution.

%% file: sections/agent.tex
\section{Part 2: Tool ablation---setup}
\label{sec:agent}

\subsection{Task and evaluation subset}
The agent solves \emph{localization only}: it reads the issue and must call \submit\ once with a ranked list of up to ten locations, each \texttt{path.py:QualifiedName} or \texttt{path.py} for module-level code. No patch is written. We draw a fixed, seeded subset of 160 tasks, 40 from each of the four hop buckets, from the 468 bucketed tasks; 156 of them have at least one function-level gold target (the other four are module-level patches and are scored on file accuracy only). The subset spans eleven repositories (Django 80, SymPy 23, Sphinx 17, Matplotlib 13, scikit-learn 7, xarray 6, Astropy 5, pytest 4, Requests 3, seaborn 1, pylint 1) and has a mean of 1.89 function-level gold targets per task (median 1, maximum 15). The budget experiment (\run{Budget}) uses 20 tasks per bucket (80 tasks, 78 with function gold) drawn from the same seed.

\subsection{Tools and arms}
\label{sec:tools_def}
Every arm has the three text tools; the structured arms add one or two more. Tool outputs are lists of lines; each line carries the set of graph nodes it exhibits, which is what makes the trace analysis of \cref{sec:b5} exact.
\begin{itemize}
\item \grep\texttt{(pattern, path\_prefix)}: regular-expression search over all \texttt{.py} files, up to 40 \texttt{path:line: text} matches; a line exhibits the innermost definition containing it.
\item \readfile\texttt{(path, start, end)}: numbered lines, at most 120 per call; each line exhibits its innermost definition.
\item \listdir\texttt{(path)}: up to 100 children of a directory; exhibits nothing.
\item \semsearch\texttt{(query)} (arms \E, \ET): the ten function or class nodes whose text is closest to the query under Qwen3-Embedding-0.6B \citep{qwen3embedding} (last-token pooling, L2-normalised, index text = node id plus the first 40 lines of the definition; query prefixed with the retrieval instruction ``Given a description of a software issue, retrieve the code that is relevant''). Each hit is one line, \texttt{id signature}, and exhibits that node.
\item \findentity\texttt{(name)} (arms \T, \ET): up to 20 nodes whose qualified name equals, ends with \texttt{.name}, or whose short name equals the query; one line per hit with its signature.
\item \graphnb\texttt{(entity, hops)} (arms \T, \ET): breadth-first expansion of the graph of \cref{sec:graph} from a node id (or a unique suffix of one) for one or two hops, in both directions, one line per edge of the form \mbox{\texttt{[hop h] u --calls--> v signature}}, truncated at 50 lines (in practice the 4{,}000-character observation budget of \cref{sec:harness} binds first). Relations are rendered as \emph{contains / calls / imports / inherits} and their inverses.
\item \submit\texttt{(locations)}: ends the episode.
\end{itemize}
Arms: $\G=\{\grep,\readfile,\listdir\}$; $\E=\G\cup\{\semsearch\}$; $\T=\G\cup\{\findentity,\graphnb\}$; $\ET=\E\cup\T$. The system prompt and tool schemas are reproduced in \cref{sec:app_prompts}.

\subsection{Harness}
\label{sec:harness}
Models \citep{qwen3} are served with vLLM \citep{vllm} through the OpenAI chat-completions interface with native tool calling (hermes parser for Qwen, gemma4 parser for Gemma). Unless stated, decoding is greedy (temperature 0), the step cap is 30 tool calls, and the context window is 32k tokens. Qwen3-8B and Gemma are run with thinking disabled. Two harness rules were found necessary in a pilot and are used in every reported run:
\begin{enumerate}
\item \emph{Forced tool call after a bare turn.} If a turn contains no tool call, the harness replies ``Use a tool, or call submit with your ranked locations'' and sets \texttt{tool\_choice=required} for the next turn. Without this, greedy decoding frequently repeats the same tool-less message until the step cap; the first full run without the rule (\run{Abl-4B-prefix}) is reported in \cref{tab:prefix} for completeness and is otherwise not used.
\item \emph{Observation budget.} Tool output is truncated at 4{,}000 characters. A node counts as \emph{revealed} only if it is on a line the model actually received; truncated lines reveal nothing.
\end{enumerate}
Every step is logged with the tool, its arguments, the output as shown, the revealed nodes, the revealed \emph{gold} nodes, token counts and wall-clock time, and the complete message list is saved as a trace. Invalid JSON arguments are replaced by an error observation; an exception from the server (context overflow, timeout) ends the episode. Options used by individual experiments: \texttt{--temperature}/\texttt{--seed} (\run{Seeds}), \texttt{--force-first} (\run{ForceTool}), \texttt{--max-steps} (\run{Budget}), \texttt{--embedder potion} (\run{WeakEmb}).

\subsection{Metrics}
\label{sec:metrics}
Let $S=(s_1,\dots,s_n)$ be the submitted list, $S^{\mathrm{f}}$ its function-level entries (those containing ``:'') in order, $G^{\mathrm{f}}$ the function-level gold targets and $F$ the set of gold files. A submitted string $s$ matches a gold $g$ if $s=g$, or one is a qualified-name prefix of the other (so submitting a class counts for a method of it and vice versa). Then
\begin{align}
\text{\facc} &= \mathbf{1}\Big[\forall g\in G^{\mathrm{f}}\ \exists s\in S^{\mathrm{f}}_{1:10}:\ \mathrm{match}(s,g)\Big],\\
\text{function recall@10} &= \frac{1}{|G^{\mathrm{f}}|}\sum_{g\in G^{\mathrm{f}}}\mathbf{1}\big[\exists s\in S^{\mathrm{f}}_{1:10}:\mathrm{match}(s,g)\big],\\
\text{file Acc@5} &= \mathbf{1}\big[F\subseteq \mathrm{files}(S)_{1:5}\big],\qquad
\text{gold revealed} = \mathbf{1}\big[\exists\,\text{step }t:\ R_t\cap G^{\mathrm{f}}\neq\varnothing\big],
\end{align}
where $R_t$ is the set of nodes revealed at step $t$ and $\mathrm{files}(S)$ the deduplicated file prefixes of $S$. We also record the first step at which a gold node is revealed, the submit rate, the number of steps and total tokens. \facc\ is the primary metric: it requires \emph{every} gold function to be in the top ten, which is the quantity a repair stage needs.

\subsection{Experiments}
\label{sec:experiments}
\cref{tab:expts} lists the runs and the identifiers used for them in the rest of the paper. The primary comparison is one greedy run of all four arms on the 160 tasks for Qwen3-4B-Instruct-2507 (\run{Abl-4B}), Qwen3-8B (\run{Abl-8B}), and Gemma~4 E4B-it (\run{Abl-Gemma}). The Gemma runs cover all 500 tasks and are reported on the 160-task subset unless stated. \run{Abl-32B} runs the four arms with Qwen3-32B (the official FP8 checkpoint, non-thinking) on all 500 tasks, and \run{Abl-4B-500} and \run{Abl-8B-500} complete the Qwen3-4B and Qwen3-8B greedy runs on the 340 tasks outside the subset, so that the three Qwen sizes can be compared on the full benchmark. \run{Seeds} adds two runs at temperature 0.7 with seeds 1 and 2 for Qwen3-4B, Qwen3-8B and Gemma~4 E4B, so that every comparison for those three models can be made on three runs. \run{Traces} is a trace analysis of those runs. \run{WeakEmb} replaces the embedder in arm \E\ with a weak static model (model2vec \texttt{potion-base-8M} \citep{model2vec}, mean of static token vectors, 256-d, no instruction prefix) for Qwen3-8B. \run{ForceTool} forces the first tool call of arm \E\ to be \semsearch\ and the first two calls of arm \T\ to be \findentity\ then \graphnb, with a 256-token cap and one retry on the forced call. \run{Budget} raises the step cap to 50 and the context window to 40{,}960 tokens for arms \G\ and \ET\ of Qwen3-8B on 80 tasks. \run{Repair} feeds each \run{Abl-8B} episode's localizations to a repair model and scores the patches with the official SWE-bench harness (\cref{sec:b8}). \run{ScaleCurve} reruns the four arms on all 500 tasks for six Qwen3 sizes (0.6B--32B, greedy and four seeds at $T{=}0.7$), Qwen3-14B-FP8 (greedy and two seeds) and Gemma~4 E2B, E4B and 12B, on a newer build of the same harness (\cref{sec:scalecurve}).

\begin{table}[t]
\centering\footnotesize\setlength{\tabcolsep}{4pt}
\caption{\textbf{Agent runs and their identifiers.} All runs use the 160-task depth-balanced subset unless noted; ``4 arms'' is \G, \E, \T, \ET. The L4 is a GCP \texttt{g2-standard-16} VM; the A100s (40 or 80\,GB) are Colab instances.}
\label{tab:expts}
\begin{tabularx}{\linewidth}{llXl}
\toprule
ID & Model & Setting & Hardware\\
\midrule
\run{Abl-4B} & Qwen3-4B-Instruct-2507 & tool ablation: 4 arms, $T{=}0$ & L4\\
\run{Abl-8B} & Qwen3-8B (non-thinking) & tool ablation: 4 arms, $T{=}0$ & A100\\
\run{Abl-Gemma} & Gemma 4 E4B-it & tool ablation: 4 arms, $T{=}0$, gemma4 tool parser; all 500 tasks & A100\\
\run{Abl-32B} & Qwen3-32B (FP8, non-thinking) & tool ablation: 4 arms, $T{=}0$; all 500 tasks & A100 80GB\\
\run{Abl-4B-500}, \run{Abl-8B-500} & Qwen3-4B, Qwen3-8B & 4 arms, $T{=}0$, the 340 tasks outside the subset (with \run{Abl-4B}/\run{Abl-8B}: all 500) & A100 80GB\\
\run{Seeds} & Qwen3-4B, Qwen3-8B, Gemma 4 E4B & run-to-run variance: 4 arms, $T{=}0.7$, seeds 1 and 2 (Gemma: all 500 tasks) & L4 / A100\\
\run{Traces} & --- & trace analysis of \run{Abl-4B}, \run{Abl-8B}, \run{Seeds} & CPU\\
\run{WeakEmb} & Qwen3-8B & retrieval quality: arm \E\ with potion-base-8M instead of Qwen3-Embedding-0.6B & A100\\
\run{ForceTool} & Qwen3-4B, Qwen3-8B & forced use: arms \E, \T; first call(s) forced to the structured tool & L4 / A100\\
\run{Budget} & Qwen3-8B & step/context budget: arms \G, \ET; 80 tasks; 50 steps; 40{,}960-token context & A100\\
\run{Repair} & Qwen3-8B (repair) & downstream repair: top-5 locations of each \run{Abl-8B} episode; official harness & A100 + L4\\
\run{ScaleCurve} & Qwen3 0.6B--32B, Gemma 4 & scale curve: 4 arms, all 500 tasks; six Qwen3 sizes (0.6B, 1.7B, 4B hybrid, 8B, 14B, 32B-FP8) at $T{=}0$ and $T{=}0.7$ seeds 1--4; Qwen3-14B-FP8 and Gemma 4 E2B/E4B at $T{=}0$ and seeds 1--2; Gemma 4 12B at $T{=}0$ only (partial) & A100 80GB\\
\bottomrule
\end{tabularx}

\end{table}

\subsection{Statistics}
\label{sec:stats}
Within a run, arms are compared on the same tasks with an exact McNemar test on per-task \facc\ (the binomial test on the discordant pairs). Across the three Qwen runs we average each task's outcome over runs and report the mean and a 95\% percentile bootstrap interval over tasks (5{,}000 resamples) for each arm and for each paired difference. The depth analysis uses the GEE model of \cref{eq:gee}. We do not correct for multiple comparisons; the number of tests is modest and we report all of them.

%% file: sections/results_tools.tex
\section{Part 2: Tool ablation---results}
\label{sec:tools}

\subsection{What each tool set contributes (\run{Abl-4B}, \run{Abl-8B}, \run{Abl-Gemma})}
\label{sec:b1}

\begin{table}[t]
\centering\small
\caption{\textbf{Tool ablation, primary greedy runs} (160 tasks; \facc\ and recall over the 156 tasks with function-level gold). Steps and submit rate are per episode. The last columns give exact McNemar $p$-values against \G\ (discordant counts in parentheses, arm-only wins / \G-only wins) on \facc.}
\label{tab:main}
\resizebox{\linewidth}{!}{%
\begin{tabular}{llcccccl}
\toprule
Model & Arm & \facc\ & func.\ recall@10 & file Acc@5 & steps & submit & vs.\ \G\ (wins)\\
\midrule
\multirow{4}{*}{Qwen3-4B (\run{Abl-4B})} & \G & 24.4 & 28.1 & 53.1 & 7.8 & 0.975 & ---\\
 & \E & 23.1 & 27.6 & 56.2 & 8.8 & 0.956 & $p=0.82$ (9/11)\\
 & \T & \textbf{30.8} & \textbf{35.2} & 56.9 & 8.5 & 0.994 & $p=0.099$ (20/10)\\
 & \ET & 29.5 & 33.2 & 56.2 & 8.9 & 0.988 & $p=0.18$ (18/10)\\
\midrule
\multirow{4}{*}{Qwen3-8B (\run{Abl-8B})} & \G & 12.8 & 14.6 & 33.1 & 16.6 & 0.619 & ---\\
 & \E & 26.3 & 29.5 & 51.2 & 11.9 & 0.819 & $p=4.9{\times}10^{-5}$ (24/3)\\
 & \T & 23.7 & 27.1 & 43.1 & 13.7 & 0.781 & $p=9.1{\times}10^{-4}$ (21/4)\\
 & \ET & \textbf{29.5} & \textbf{34.1} & \textbf{53.1} & 11.5 & 0.850 & $p=1.3{\times}10^{-5}$ (31/5)\\
\midrule
\multirow{4}{*}{Gemma 4 E4B} & \G & 9.6 & 11.4 & 53.8 & 7.4 & 0.969 & ---\\
 & \E & 21.8 & 25.1 & \textbf{60.0} & 6.4 & 0.975 & $p=5.5{\times}10^{-4}$ (24/5)\\
 & \T & 21.8 & 24.1 & 53.1 & 8.5 & 0.963 & $p=5.5{\times}10^{-4}$ (24/5)\\
 & \ET & \textbf{23.7} & \textbf{27.0} & 59.4 & 7.5 & 0.963 & $p=1.1{\times}10^{-4}$ (27/5)\\
\midrule
\multirow{4}{*}{Qwen3-32B (\run{Abl-32B}, subset)} & \G & 26.9 & 31.1 & 59.4 & 5.7 & 0.988 & ---\\
 & \E & \textbf{32.1} & \textbf{36.6} & \textbf{65.6} & 5.5 & 0.981 & $p=0.12$ (14/6)\\
 & \T & 28.2 & 33.1 & 65.0 & 7.4 & 0.975 & $p=0.84$ (13/11)\\
 & \ET & 28.8 & 33.5 & 61.9 & 6.6 & 0.975 & $p=0.63$ (10/7)\\
\bottomrule
\end{tabular}}

\end{table}

\begin{table}[t]
\centering\small
\caption{\textbf{Three-run estimates} (greedy run plus two seeds at $T{=}0.7$; Gemma's runs restricted to the subset). Each task's \facc\ is averaged over runs; entries are means over 156 tasks with 95\% percentile-bootstrap intervals over tasks. Differences are paired.}
\label{tab:pooled}
\resizebox{\linewidth}{!}{%
\begin{tabular}{lcccccc}
\toprule
 & \multicolumn{2}{c}{Qwen3-4B} & \multicolumn{2}{c}{Qwen3-8B} & \multicolumn{2}{c}{Gemma 4 E4B}\\
\cmidrule(lr){2-3}\cmidrule(lr){4-5}\cmidrule(lr){6-7}
Arm & mean & 95\% CI & mean & 95\% CI & mean & 95\% CI\\
\midrule
\G & 22.6 & \ci{16.7}{28.8} & 13.0 & \ci{8.5}{17.9} & 13.5 & \ci{9.6}{17.7}\\
\E & 21.6 & \ci{16.2}{27.4} & 24.6 & \ci{18.6}{30.6} & 25.2 & \ci{19.7}{31.2}\\
\T & 30.8 & \ci{24.1}{37.6} & 24.1 & \ci{17.9}{30.3} & 22.4 & \ci{16.9}{28.2}\\
\ET & 28.0 & \ci{22.0}{34.2} & 26.5 & \ci{20.5}{32.9} & 27.1 & \ci{21.4}{33.1}\\
\midrule
\E$-$\G & $-1.1$ & \ci{-4.9}{+2.6} & $+11.5$ & \ci{+7.1}{+16.2} & $+11.8$ & \ci{+7.5}{+16.5}\\
\T$-$\G & $+8.1$ & \ci{+3.0}{+13.5} & $+11.1$ & \ci{+6.8}{+15.8} & $+9.0$ & \ci{+4.7}{+13.5}\\
\ET$-$\G & $+5.3$ & \ci{+0.0}{+10.5} & $+13.5$ & \ci{+8.5}{+18.6} & $+13.7$ & \ci{+9.0}{+18.6}\\
\T$-$\E & $+9.2$ & \ci{+3.8}{+15.0} & $-0.4$ & \ci{-4.7}{+3.8} & $-2.8$ & \ci{-7.3}{+1.7}\\
\ET$-$\E & $+6.4$ & \ci{+1.3}{+11.5} & $+1.9$ & \ci{-2.4}{+6.4} & $+1.9$ & \ci{-2.6}{+6.2}\\
\ET$-$\T & $-2.8$ & \ci{-6.8}{+1.1} & $+2.4$ & \ci{-0.9}{+6.0} & $+4.7$ & \ci{+0.4}{+9.2}\\
\bottomrule
\end{tabular}}
\end{table}

\cref{tab:main} gives the primary greedy runs and \cref{tab:pooled} the three-run estimates. Three patterns hold.

\textbf{Grep alone is a weak baseline for the 8B and Gemma models, and any structured tool roughly doubles it.} Qwen3-8B with \G\ reaches 12.8\% \facc\ (13.0\% over three runs); \E, \T\ and \ET\ reach 26.3, 23.7 and 29.5\% (24.6, 24.1, 26.5\% over three runs), all differences from \G\ significant at $p<10^{-3}$ in the greedy run and with bootstrap intervals that exclude zero by a wide margin. Gemma~4 E4B behaves the same way (9.6 $\to$ 21.8 / 21.8 / 23.7\% in the greedy run; 13.5 $\to$ 25.2 / 22.4 / 27.1\% over three runs). File-level accuracy tells the same story for Qwen3-8B (33.1 $\to$ 51.2 / 43.1 / 53.1\%) but not for Gemma~4 E4B, whose file Acc@5 is already 53.8\% with \grep\ and does not change significantly: for Gemma the tools improve the \emph{function} the agent names inside a file it already found.

\textbf{For Qwen3-4B only the graph tools help.} \G\ is much stronger for the 4B model (24.4\%) than for the 8B model, and embedding search adds nothing: $-1.1$ points over three runs, $\ci{-4.9}{+2.6}$. The graph tools add $+8.1$ points $\ci{+3.0}{+13.5}$ (significant at $p\le0.035$ in each of the two sampled runs and $p=0.099$ in the greedy run), and giving the 4B model all tools is slightly \emph{worse} than the graph tools alone ($-2.8$, $\ci{-6.8}{+1.3}$). Section \ref{sec:b5} shows why: the 4B model calls \semsearch\ in only 43\% of \E\ episodes and 14\% of \ET\ episodes.

\textbf{Combining tools helps the 8B and Gemma models modestly.} For Qwen3-8B, \ET\ beats \T\ in the greedy run ($p=0.049$, 13/4) and by $+2.4$ $\ci{-0.9}{+6.0}$ over three runs; \ET\ versus \E\ is $+1.9$ $\ci{-2.4}{+6.4}$. For Gemma~4 E4B the greedy run has the three structured arms within 2 points of one another, but over three runs \ET\ leads \T\ by $+4.7$ $\ci{+0.4}{+9.2}$ and \E\ by $+1.9$ $\ci{-2.6}{+6.2}$. The two tools reveal gold code through different routes (\cref{sec:b5}) and their union is larger than either.

\begin{figure}[t]
\centering
\includegraphics[width=\linewidth]{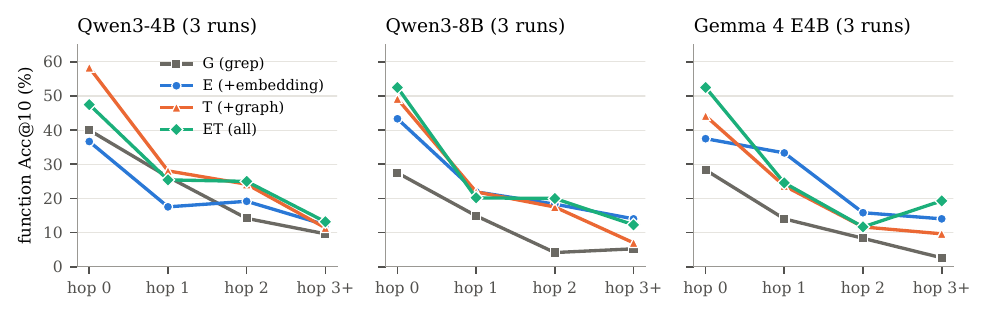}
\caption{\textbf{\facc\ by hop bucket and tool set.} All panels pool three runs (one greedy, two sampled). Each bucket has 40 tasks (38--40 with function-level gold). The advantage of the graph tools (\T) over \G\ is largest at hop 0 (16--22 points) and shrinks to 2 points by hop 3+ for the Qwen models (7 for Gemma, whose \G\ arm falls to 2.6\%); the embedding arm (\E) has the flattest slope.}
\label{fig:bybucket}
\end{figure}

\paragraph{Accuracy by depth.}
\figref{fig:bybucket} breaks the three-run estimates down by bucket (per-run numbers are in \cref{tab:bybucket_runs}). Every arm of every model falls with depth, from 28--58\% at hop 0 to 3--19\% at hop 3+. The shape of the fall differs by tool. For Qwen3-4B, \T\ leads at hop 0 by 18 points (58.3 vs 40.0\%) and by 10 at hop 2, but by 2 at hop 3+ (11.4 vs 9.6\%). For Qwen3-8B, \T\ leads \G\ by 22 points at hop 0 and by \emph{2} at hop 3+ (7.0 vs 5.3\%), while \E\ leads by 16 at hop 0 and still by 9 at hop 3+ (14.0\%). For Gemma~4 E4B, \G\ falls to 2.6\% at hop 3+ while \E\ and \ET\ keep 14.0 and 19.3\%.

The GEE interaction model (\cref{tab:gee_extra}) makes this precise. Fitting \cref{eq:gee} separately per arm, the odds ratio per hop is 0.47 (\G), 0.59 (\E), 0.34 (\T) and 0.52 (\ET) for Qwen3-4B; 0.41, 0.55, 0.35 and 0.43 for Qwen3-8B over three runs; 0.39, 0.57, 0.45 and 0.48 for Gemma~4 E4B over three runs. In the joint model, the $\hop\times\E$ interaction is 1.24 ($p=0.048$) for Qwen3-4B and 1.30 ($p=0.098$) for Gemma~4 E4B, \ie\ \E\ degrades less steeply than \G; the $\hop\times\T$ interaction is 0.84, 0.89 and 1.01 (all $p>0.3$), \ie\ the graph arm's slope is not distinguishable from \grep's and, in the Qwen models, is point-estimated as steeper. This is unlike the pattern LocAgent reports for its graph agent with larger models \citep{locagent}; with 4--8B models the benefit of graph tools is a shallow-depth benefit, and \cref{sec:scale,sec:scalecurve} show that the benefit itself shrinks with scale. \cref{sec:b5} shows the mechanism.

\subsection{Run-to-run variance (\run{Seeds})}
\label{sec:b4}
\cref{tab:b4} lists \facc\ per run. The arm ordering is stable: in all three Qwen3-4B runs \T\ is best (30.8, 30.8, 30.8\%) and \E\ is within 2 points of \G\ (above it once); in all three Qwen3-8B runs \G\ is far below the others (12.8, 12.2, 14.1\%) and the three structured arms are within 5 points of one another with no consistent winner (the greedy run favours \ET, seed 1 favours \ET, seed 2 favours \E); in all three Gemma runs \G\ is far below the others (9.6, 14.7, 16.0\%) and \ET\ is best or tied for best. The between-run standard deviation of an arm's mean is 0--3.4 points, which is why single-run comparisons among the structured arms of the 8B model should not be over-read, and why we report three-run intervals.

\begin{table}[t]
\centering\small
\caption{\textbf{\run{Seeds}: \facc\ (\%) per run on the 160-task subset.} The greedy run is \run{Abl-4B}/\run{Abl-8B}/\run{Abl-Gemma}; the others sample at $T{=}0.7$ with the given seed. SD is the between-run standard deviation of the arm mean.}
\label{tab:b4}
\setlength{\tabcolsep}{3.5pt}
\begin{tabular}{lcccc@{\hskip 1em}cccc@{\hskip 1em}cccc}
\toprule
 & \multicolumn{4}{c}{Qwen3-4B} & \multicolumn{4}{c}{Qwen3-8B} & \multicolumn{4}{c}{Gemma 4 E4B}\\
\cmidrule(lr){2-5}\cmidrule(lr){6-9}\cmidrule(lr){10-13}
Run & \G & \E & \T & \ET & \G & \E & \T & \ET & \G & \E & \T & \ET\\
\midrule
greedy & 24.4 & 23.1 & 30.8 & 29.5 & 12.8 & 26.3 & 23.7 & 29.5 & 9.6 & 21.8 & 21.8 & 23.7\\
$T{=}0.7$, seed 1 & 23.7 & 19.9 & 30.8 & 29.5 & 12.2 & 21.8 & 23.7 & 26.3 & 14.7 & 28.2 & 21.2 & 29.5\\
$T{=}0.7$, seed 2 & 19.9 & 21.8 & 30.8 & 25.0 & 14.1 & 25.6 & 25.0 & 23.7 & 16.0 & 25.6 & 24.4 & 28.2\\
\midrule
SD & 2.4 & 1.6 & 0.0 & 2.6 & 1.0 & 2.4 & 0.7 & 2.9 & 3.4 & 3.2 & 1.7 & 3.0\\
\bottomrule
\end{tabular}
\end{table}

\subsection{Scale and the full benchmark (\run{Abl-32B}, \run{Abl-4B-500}, \run{Abl-8B-500})}
\label{sec:scale}
Three further runs ask whether the picture changes with scale and with the full task set. \run{Abl-32B} runs the four arms with Qwen3-32B on all 500 tasks, and \run{Abl-4B-500} and \run{Abl-8B-500} complete the Qwen3-4B and Qwen3-8B greedy runs on the 340 tasks outside the subset; with Gemma~4 E4B's greedy run, which also covers all 500 tasks, \cref{tab:full500} compares four models on the whole benchmark (485 tasks with function-level gold). The full set is easier than the depth-balanced subset, since 186 of its tasks are at hop 0.

\begin{table}[t]
\centering\small
\caption{\textbf{All 500 tasks: \facc\ overall and by bucket, file Acc@5, steps and submit rate}, one greedy run per model; McNemar versus \G\ over the 485 tasks with function-level gold (wins/losses).}
\label{tab:full500}
\setlength{\tabcolsep}{3.5pt}\footnotesize
\resizebox{\linewidth}{!}{%
\begin{tabular}{llcc@{\hskip 1em}cccc@{\hskip 1em}ccl}
\toprule
 & & \multicolumn{2}{c}{overall} & \multicolumn{4}{c}{\facc\ by bucket} & & &\\
\cmidrule(lr){3-4}\cmidrule(lr){5-8}
Model & Arm & \facc & file Acc@5 & hop 0 & hop 1 & hop 2 & hop 3+ & steps & submit & vs.\ \G\\
\midrule
\multirow{4}{*}{Qwen3-4B} & \G & 26.8 & 61.2 & 36.8 & 33.3 & 13.2 & 6.8 & 7.6 & 0.982 & ---\\
 & \E & 29.1 & 63.0 & 38.4 & 34.1 & 10.5 & 20.5 & 8.0 & 0.970 & $p=0.24$ (41/30)\\
 & \T & \textbf{37.1} & \textbf{65.4} & 53.5 & 39.8 & 19.7 & 15.1 & 8.4 & 0.994 & $p=5.6{\times}10^{-7}$ (75/25)\\
 & \ET & 36.5 & 64.8 & 53.0 & 40.7 & 21.1 & 11.0 & 8.8 & 0.988 & $p=2.5{\times}10^{-6}$ (73/26)\\
\midrule
\multirow{4}{*}{Qwen3-8B} & \G & 16.7 & 41.8 & 24.3 & 25.2 & 3.9 & 2.7 & 15.8 & 0.638 & ---\\
 & \E & 30.3 & 55.2 & 41.6 & 33.3 & 17.1 & 13.7 & 12.3 & 0.794 & $p=2.1{\times}10^{-12}$ (80/14)\\
 & \T & 27.4 & 48.4 & 45.4 & 26.0 & 13.2 & 6.8 & 14.4 & 0.748 & $p=9.4{\times}10^{-8}$ (74/22)\\
 & \ET & \textbf{33.2} & \textbf{58.6} & 48.6 & 34.1 & 21.1 & 13.7 & 11.5 & 0.854 & $p=1.0{\times}10^{-15}$ (94/14)\\
\midrule
\multirow{4}{*}{Gemma 4 E4B} & \G & 16.7 & 64.2 & 23.8 & 19.5 & 7.9 & 1.4 & 7.1 & 0.984 & ---\\
 & \E & 28.5 & \textbf{66.4} & 34.1 & 36.6 & 14.5 & 16.4 & 6.3 & 0.984 & $p=6.9{\times}10^{-9}$ (78/21)\\
 & \T & 26.8 & 64.0 & 37.8 & 32.5 & 11.8 & 8.2 & 8.4 & 0.972 & $p=6.4{\times}10^{-7}$ (73/24)\\
 & \ET & \textbf{29.7} & 65.8 & 41.6 & 33.3 & 11.8 & 13.7 & 7.3 & 0.976 & $p=6.4{\times}10^{-10}$ (85/22)\\
\midrule
\multirow{4}{*}{Qwen3-32B} & \G & 32.0 & 68.4 & 42.2 & 41.5 & 11.8 & 16.4 & 5.2 & 0.988 & ---\\
 & \E & 35.7 & \textbf{71.6} & 44.3 & 43.1 & 19.7 & 21.9 & 5.2 & 0.988 & $p=0.027$ (39/21)\\
 & \T & 34.0 & 70.4 & 46.5 & 40.7 & 17.1 & 15.1 & 7.3 & 0.970 & $p=0.27$ (38/28)\\
 & \ET & \textbf{35.9} & 70.2 & 49.2 & 44.7 & 10.5 & 19.2 & 6.3 & 0.984 & $p=0.027$ (43/24)\\
\bottomrule
\end{tabular}}
\end{table}

\textbf{The structured tools' gain shrinks from 10--17 points at 4--8B to 2--4 points at 32B.} On the full benchmark, Qwen3-4B goes from 26.8\% with \grep\ alone to 37.1 / 36.5\% with the graph tools (\T/\ET, $p<10^{-5}$) and, as on the subset, gains nothing from embedding search alone (29.1\%, $p=0.24$); Qwen3-8B goes from 16.7\% to 30.3 / 27.4 / 33.2\% with \E/\T/\ET, and Gemma~4 E4B from 16.7 to 28.5 / 26.8 / 29.7\%, every difference from \G\ significant at $p<10^{-6}$. Qwen3-32B goes from 32.0 to 35.7 / 34.0 / 35.9\%: \E\ and \ET\ add 3.7--3.9 points ($p=0.027$ each) and \T\ adds 2.0 ($p=0.27$). A joint model over the three Qwen sizes on the same 500 tasks makes the trend a test: the $\mathrm{arm}\times\log(\text{size})$ interaction is 0.76 $\ci{0.66}{0.86}$ for \T\ and 0.78 $\ci{0.68}{0.89}$ for \ET\ (both $p<10^{-3}$), \ie\ each doubling of parameters removes about a fifth of the graph tools' odds advantage, while the $\hop\times\log(\text{size})$ interaction is 1.05 $\ci{0.97}{1.14}$ ($p=0.20$): the depth gradient does not change with scale. The 32B model with \grep\ alone (32.0\%) matches the 8B model with every tool (33.2\%), and its file-level accuracy with \grep\ alone (68.4\%) exceeds the 8B model's best (58.6\%). On the 160-task subset the picture is the same with wider intervals: 26.9 $\to$ 32.1 / 28.2 / 28.8\% (\cref{tab:main}; \E\ versus \G\ $p=0.12$).

\textbf{The larger model does not use the structured tools more; it uses them less.} Qwen3-32B calls \findentity\ in 39\% of its \T\ episodes and 44\% of its \ET\ episodes, \graphnb\ in 31\% and 15\%, and \semsearch\ in 34\% of \E\ and 8\% of \ET\ episodes, against 94\%, 86\% and 85\% for Qwen3-8B and 98\%, 87\% and 40\% for Qwen3-4B on the same 500 tasks (\cref{tab:b5}). It never requests a two-hop neighbourhood (0 of 263 \graphnb\ calls ask for \texttt{hops=2}) and chains expansions in 0.7\% of \T/\ET\ episodes. What it does instead is use \grep\ and \readfile\ well: 1.8 \grep\ and 2.4 \readfile\ calls per \G\ episode, 5.2 steps, 98.8\% submission, 1\% at the cap, and a \grep-only discovery rate of 66\% with 45\% conversion, against 55\% and 30\% for the 8B model on the same tasks. The gain that the structured tools provide at 4--8B---exact-name resolution and one expansion---is something the 32B model largely obtains from the plain file tools by itself, so the tools have less to add; \cref{sec:scalecurve} refines this with a 14B model whose \grep\ arm is as accurate as the 32B model's but which still calls the tools and still gains from them. In arm \T\ at hop 0, \findentity\ is the first tool to reveal the gold in 18\% of 32B episodes against 46\% for 8B; \grep\ is first in 63\% (\cref{tab:firsttool}).

\textbf{Depth remains the dominant difficulty axis at 32B.} With controls, each hop multiplies the 32B model's odds of correct localization by 0.552 $\ci{0.453}{0.671}$ ($p=2.7{\times}10^{-9}$; 457 tasks, 1{,}828 episodes), against 0.469 $\ci{0.393}{0.561}$ for Qwen3-4B and 0.478 $\ci{0.396}{0.578}$ for Qwen3-8B on the same 500 tasks (\cref{tab:c2}). Per arm the 32B odds ratios are 0.56 (\G), 0.62 (\E), 0.52 (\T) and 0.49 (\ET), and no interaction with depth is significant (all $p\ge0.18$). The shape of the fall changes: the 32B \G\ arm is flat across hop 0 and hop 1 (42.2 and 41.5\%) and halves at hop 2 (11.8\%), so what a larger model gains is the one-edge step that the graph tool gives the small models, not the deeper ones. These runs are single greedy runs (the 32B model served from its FP8 checkpoint), so the arm-versus-arm differences at 32B should be read with the $\pm$3-point resolution of one run on 485 tasks; \cref{sec:scalecurve} reruns them with four sampled decodings each and an FP8 control.

\input{sections/scale}

\subsection{Trace analysis: discovery, conversion and which tool finds the gold (\run{Traces})}
\label{sec:b5}

\begin{table}[t]
\centering\small
\caption{\textbf{\run{Traces}: episode outcomes and tool use}, pooled over three runs per model (156 tasks with function gold per run). \emph{saw} is the fraction of episodes in which some gold function was revealed; \emph{conv.} is $\Pr(\text{correct}\mid\text{saw})$; \emph{cap} is the fraction that hit the 30-step cap. Right block: calls per episode (ss = \semsearch, fe = \findentity, gn = \graphnb) and the fraction of episodes with at least one call to the structured tool.}
\label{tab:b5}
\setlength{\tabcolsep}{3pt}\footnotesize
\resizebox{\linewidth}{!}{%
\begin{tabular}{llcccccc@{\hskip 1.2em}ccccc@{\hskip 1.2em}cc}
\toprule
 & & & & & & & & \multicolumn{5}{c}{calls / episode} & \multicolumn{2}{c}{eps.\ using}\\
\cmidrule(lr){9-13}\cmidrule(lr){14-15}
Model & arm & acc & saw & conv. & submit & cap & steps & \grep & \readfile & ss & fe & gn & ss & fe/gn\\
\midrule
Qwen3-4B & \G & .226 & .562 & .380 & .983 & .017 & 8.0 & 3.8 & 1.9 & --- & --- & --- & --- & ---\\
 & \E & .216 & .588 & .338 & .951 & .045 & 8.8 & 3.2 & 2.6 & 0.59 & --- & --- & .43 & ---\\
 & \T & .308 & .660 & .456 & .989 & .004 & 8.6 & 2.6 & 1.3 & --- & 1.54 & 1.37 & --- & .98/.90\\
 & \ET & .280 & .658 & .416 & .974 & .024 & 9.2 & 2.6 & 1.6 & 0.23 & 1.73 & 1.18 & .14 & .98/.87\\
\midrule
Qwen3-8B & \G & .130 & .481 & .271 & .596 & .402 & 17.1 & 10.0 & 3.6 & --- & --- & --- & --- & ---\\
 & \E & .246 & .596 & .401 & .784 & .207 & 12.9 & 7.3 & 2.8 & 1.38 & --- & --- & .85 & ---\\
 & \T & .241 & .626 & .386 & .761 & .226 & 14.1 & 4.9 & 2.8 & --- & 1.69 & 2.74 & --- & .94/.86\\
 & \ET & .265 & .654 & .399 & .821 & .165 & 12.2 & 5.9 & 2.0 & 0.53 & 0.89 & 1.51 & .52 & .72/.78\\
\midrule
Gemma 4 E4B & \G & .135 & .526 & .211 & .981 & .019 & 7.5 & 1.8 & 2.2 & --- & --- & --- & --- & ---\\
 & \E & .252 & .641 & .353 & .972 & .028 & 6.9 & 1.2 & 2.5 & 0.79 & --- & --- & .76 & ---\\
 & \T & .224 & .620 & .352 & .970 & .030 & 8.9 & 1.5 & 2.1 & --- & 1.07 & 0.98 & --- & .78/.74\\
 & \ET & .271 & .686 & .386 & .972 & .028 & 7.7 & 1.0 & 2.3 & 0.71 & 0.34 & 1.12 & .69 & .29/.74\\
\bottomrule
\end{tabular}}
\end{table}

Because every tool line is annotated with the nodes it exhibits, we can split each episode into \emph{discovery} (was a gold function ever on the screen?) and \emph{conversion} (given that it was, did the agent submit it?). \cref{tab:b5} reports both, with the tool usage that produced them.

\textbf{Structured tools raise discovery by 3--17 points and, except for Qwen3-4B's \E\ arm, conversion by 4--18 points.} For Qwen3-8B, the gold is revealed in 48\% of \G\ episodes and 60--65\% of \E/\T/\ET\ episodes, and conversion rises from 27\% to 39--40\%. For Qwen3-4B, \T\ raises discovery from 56 to 66\% and conversion from 38 to 46\%, whereas \E\ raises discovery slightly (59\%) but \emph{lowers} conversion (34\%): the 4B model is distracted by the extra results it does not act on. For Gemma~4 E4B both move, discovery from 53 to 62--69\% and conversion from 21 to 35--39\%: with \grep\ alone Gemma often finds the right file but names the wrong function.

\textbf{Grep-only Qwen3-8B collapses into search loops.} Its \G\ episodes average 10.0 \grep\ calls (the most frequent transition in the trace is \grep$\to$\grep, 8.5 times per episode), 40\% hit the 30-step cap, 41\% never submit, and in 10\% of episodes the gold was on screen but the agent never submitted anything. The structured tools cut the step count from 17.2 to 12.2--14.1 and the non-submission rate by half. The budget experiment (\cref{sec:bcap}) shows that this is not a budget artifact.

\begin{figure}[t]
\centering
\includegraphics[width=\linewidth]{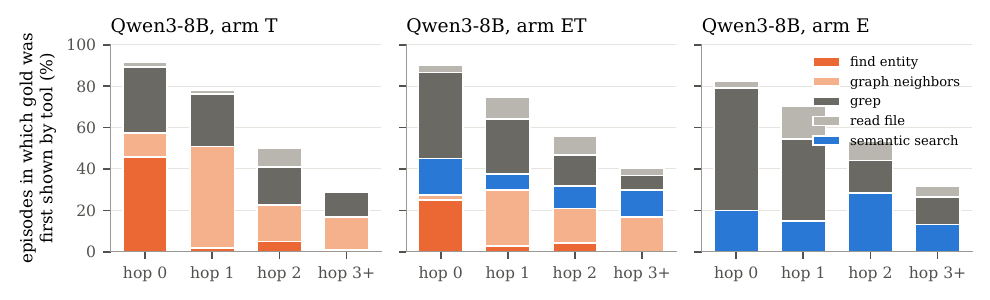}
\caption{\textbf{Which tool first shows the gold, by depth} (Qwen3-8B, three runs pooled; the 4B and Gemma panels are in \cref{tab:firsttool}). In arm \T, \findentity\ is the first tool to reveal the gold in 46\% of hop-0 episodes and \graphnb\ in 49\% of hop-1 episodes; at hop 2+ the graph tools rarely reach it and most episodes never see the gold at all. In arm \E, \semsearch\ accounts for 13--28\% of first reveals at every depth.}
\label{fig:firsttool}
\end{figure}

\textbf{The graph tools find the gold by name lookup at hop 0 and by one expansion at hop 1.} \figref{fig:firsttool} shows, for each bucket, which tool first put a gold function on the screen. In arm \T\ of Qwen3-8B, \findentity\ is the first revealing tool in 46\% of hop-0 episodes (53\% for Qwen3-4B): the issue names the target, and a name lookup returns it. At hop 1, \graphnb\ is first in 49\% (40\% for 4B): the target is one \emph{contains}/\emph{calls}/\emph{inherits} line away from a named entity. At hop 2, \graphnb\ is first in only 17\%, at hop 3+ in 16\%, and 50--70\% of episodes never see the gold. Over all steps rather than the first, \graphnb\ shows the gold at some point in 65\% of hop-1 \T\ episodes but 25\% of hop-2 and hop-3+ episodes. A two-hop neighbourhood of a mentioned node contains hop-2 targets by construction, and the agents do ask for two hops (\texttt{hops=2} in 57\% of Qwen3-4B's and 95\% of Qwen3-8B's \graphnb\ calls), but the output is cut by the 4{,}000-character observation budget in 64\% and 72\% of calls respectively (a median of 23--24 edge lines is shown, so the 50-line cap never binds), and the agents almost never \emph{chain}: in the greedy runs only 12\% (4B) and 14\% (8B) of \T/\ET\ episodes contain a \graphnb\ call whose entity was returned by the previous \graphnb\ call, and 42\% of Qwen3-8B's later \graphnb\ calls re-expand the entity it just expanded. The graph tool is used as a lookup, not as a walk. In arm \E, \semsearch\ is the first revealing tool in 13--28\% of episodes at every depth for Qwen3-8B and 24--42\% for Gemma, which is why its depth slope is flattest: semantic similarity between the issue and the target function does not depend on their graph distance.

\textbf{Graph tools also change the \emph{form} of the submission.} Of the top-5 function strings submitted by Qwen3-8B, 85\% (\T) and 84\% (\ET) are exact node ids of the snapshot, against 74\% (\E) and 47\% (\G): with \grep\ alone the model has to assemble a qualified name from file paths and \texttt{def} lines and often gets the class prefix wrong, whereas \findentity\ and \graphnb\ hand it ids verbatim. The prefix-matching in \cref{sec:metrics} forgives some of this at scoring time; the repair stage of \cref{sec:b8} does not.

\subsection{Does the embedding arm depend on retrieval quality? (\run{WeakEmb})}
\label{sec:b6}
Replacing Qwen3-Embedding-0.6B with the static potion-base-8M embedder lowers Qwen3-8B's \E\ arm from 26.3 to 21.2\% \facc\ (paired: 12 tasks lost, 4 gained, $p=0.077$) and file Acc@5 from 51.2 to 47.5\% ($p=0.39$). The agent calls \semsearch\ equally often (1.06 vs 1.00 calls per episode), but the fraction of those calls that reveal a gold function halves (54/170 = 32\% vs 25/160 = 16\%), and \semsearch\ is the first revealing tool in 11 rather than 31 episodes. Half of the \E\ arm's advantage over \G\ thus comes from the quality of the retriever and half from having \emph{any} dense retriever that occasionally lands on the right function and from the behavioural change it induces (fewer \grep\ loops: 12.3 steps versus 16.6 for \G).

\subsection{Forcing the structured tool (\run{ForceTool})}
\label{sec:b7}
If the structured tools help only when the agent chooses to call them, forcing the call might help more. It does not. Forcing \semsearch\ as the first call leaves Qwen3-8B's \E\ arm at exactly 26.3\% (11 tasks gained, 11 lost) and moves Qwen3-4B's from 23.1 to 27.6\% ($p=0.23$). Forcing \findentity\ then \graphnb\ as the first two calls moves \T\ from 23.7 to 21.8\% for Qwen3-8B ($p=0.61$) and from 30.8 to 29.5\% for Qwen3-4B ($p=0.80$). The first 4B run of this experiment had a harness bug---under constrained decoding at temperature 0 the forced \findentity\ call sometimes generated an unbounded argument string, so 42\% of \T\ episodes never submitted (19.2\%); the rerun caps the forced call at 256 tokens with one retry and is the number reported. The conclusion is that voluntary use is not the bottleneck: the 8B model already calls \semsearch\ in 85\% of its \E\ episodes and \findentity\ in 94\% of its \T\ episodes, and the information the tool returns on the first call is what it is.

\subsection{Is the grep-only collapse a budget artifact? (\run{Budget})}
\label{sec:bcap}
On 80 tasks (78 with function gold), raising the cap from 30 to 50 steps and the context from 32k to 40{,}960 tokens does not help Qwen3-8B's \G\ arm: 12.8\% \facc\ at 50 steps against 15.4\% at 30 steps on the same tasks (2 tasks lost, 0 gained), with mean steps rising from 16.1 to 21.6 and 29\% of episodes still hitting the larger cap. \ET\ likewise moves from 38.5 to 33.3\% (5 lost, 1 gained, $p=0.22$). More budget gives the looping agent more loop.

\subsection{Downstream repair (\run{Repair})}
\label{sec:b8}
Localization is only useful if a repair stage can use it. For each task and arm of \run{Abl-8B} we take the top-5 submitted strings that are exact function or class ids of the snapshot, show their source (up to 150 lines each) together with the issue to Qwen3-8B (greedy, non-thinking) with a SEARCH/REPLACE edit-block prompt (\cref{sec:app_prompts}), apply each block only if its SEARCH text occurs exactly once in the file, convert the result to a unified diff, and score it with the official SWE-bench harness 4.1.0 (Docker, prebuilt instance images). An \emph{oracle} arm shows the gold functions instead. Episodes that submitted no valid function id get no prompt and count as unresolved, so the arms are compared on all 160 tasks.

\begin{figure}[t]
\centering
\begin{minipage}[c]{0.44\linewidth}
\includegraphics[width=\linewidth]{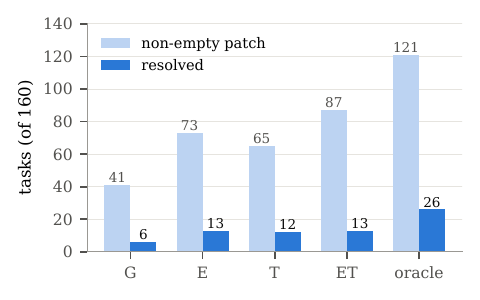}
\end{minipage}\hfill
\begin{minipage}[c]{0.55\linewidth}
\footnotesize\setlength{\tabcolsep}{3.5pt}
\resizebox{\linewidth}{!}{%
\begin{tabular}{lccccc}
\toprule
 & \G & \E & \T & \ET & oracle\\
\midrule
prompts (valid ids) & 47 & 100 & 89 & 113 & 156\\
non-empty patches & 41 & 73 & 65 & 87 & 121\\
resolved & 6 & 13 & 12 & 13 & 26\\
resolved $\mid$ loc.\ correct & 5/20 & 11/41 & 10/37 & 11/46 & ---\\
resolved $\mid$ loc.\ wrong & 1/136 & 2/115 & 2/119 & 2/110 & ---\\
\bottomrule
\end{tabular}}
\end{minipage}
\caption{\textbf{\run{Repair}: downstream repair with a fixed Qwen3-8B repair prompt} on the 160 tasks, scored with the official harness. Left: patches and resolved tasks per arm. Right: counts; ``loc.\ correct'' is \facc$=1$ in the \run{Abl-8B} episode.}
\label{fig:b8}
\end{figure}

\figref{fig:b8} summarises the result. The arms resolve 6 (\G), 13 (\E), 12 (\T) and 13 (\ET) of 160 tasks; oracle locations resolve 26. Against \G, the gains are 8/1, 7/1 and 9/2 discordant tasks ($p=0.039$, $0.070$, $0.065$); the three structured arms are indistinguishable from one another; and \ET\ is well short of the oracle (15 oracle-only tasks versus 2, $p=0.0023$). Resolution is almost entirely conditional on correct localization: among tasks the arm localized correctly, 24--27\% are resolved; among the others, 1--2\%. Resolved tasks are spread over all four buckets (for \ET: 6/2/3/2 at hop 0/1/2/3+; oracle: 5/8/7/6). Two further observations connect this to the trace analysis. First, the number of prompts---episodes whose submission contains at least one \emph{valid} id---is 47 for \G\ against 89--113 for the structured arms, which is the ``form'' effect of \cref{sec:b5} turned into a repair-stage loss. Second, the repair model is itself weak (26/156 with gold locations), so this experiment bounds what localization can contribute for a small repair model rather than measuring the ceiling of the pipeline.

%% file: sections/scale.tex
\subsection{A model-size scale curve from 0.6B to 32B (\run{ScaleCurve})}
\label{sec:scalecurve}

\begin{figure}[t]
\centering
\includegraphics[width=\linewidth]{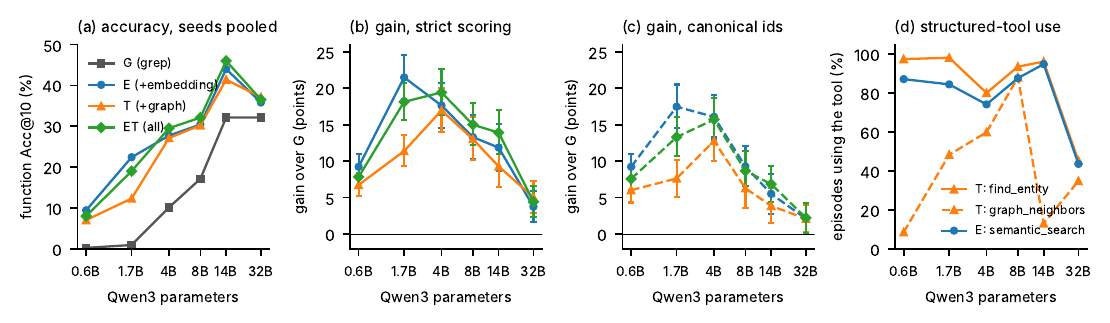}
\caption{\textbf{\run{ScaleCurve} over the six Qwen3 sizes} ($T{=}0.7$ seeds pooled, all 500 tasks). (a) Strict \facc\ per arm. (b) Gain of each structured arm over \G\ with 95\% bootstrap intervals over tasks; the gain peaks at 1.7--4B and falls to 4--5 points at 32B. (c) The same gain under canonical rescoring: from 8B up, about half of the strict gain is the form of the submission. (d) Fraction of episodes in which the agent calls the structured tool at all: \findentity\ survives to 14B, \graphnb\ is largely abandoned after 8B, and at 32B every structured tool is called in under half of the episodes.}
\label{fig:scale}
\end{figure}

The runs of \cref{sec:scale} compare three sizes with one greedy run each. \run{ScaleCurve} turns that comparison into a curve. It reruns the four arms on all 500 tasks for six Qwen3 sizes---0.6B, 1.7B, 4B, 8B, 14B and 32B (the FP8 checkpoint), all hybrid checkpoints with thinking disabled---each decoded greedily and at $T{=}0.7$ with seeds 1--4; for Qwen3-14B-FP8 (greedy and seeds 1--2) as a precision control for the 32B result; and for Gemma~4 E2B and E4B (greedy and seeds 1--2) and Gemma~4 12B (greedy only; the serving session ended after 307 tasks). The tools, limits, prompts and metrics are those of \cref{sec:agent}; the harness build is newer (vLLM 0.32.0, the path-fixed agent of \cref{sec:app_compute}) and every run of a model was served in one session. In total 79{,}228 episodes. Two things differ from the primary runs and matter for reading the results. First, the 4B model here is \texttt{Qwen/Qwen3-4B} (hybrid), not \texttt{Qwen3-4B-Instruct-2507}; we return to the difference below. Second, alongside the strict \facc\ of \cref{sec:metrics} we report a \emph{canonical} rescoring that the Colab notebooks computed with the graph cache mounted: each submitted string is repaired to a node id when the repair is unambiguous (a line number maps to the innermost definition containing it, \texttt{::} to \texttt{.}, a trailing argument list is dropped, and a bare short name is accepted when exactly one definition in the named file carries it), and the repaired list is scored as before. The difference between the two scores isolates the \emph{form} effect of \cref{sec:b5}. The planned statistic is, per model and arm, the per-task mean over the $T{=}0.7$ seeds of the arm$-$\G\ difference, with a percentile bootstrap over the 485 tasks with function-level gold, a sign-flip $p$-value, and a Holm correction over the 18 Qwen size$\times$arm tests; the greedy runs are reported alongside. The greedy runs reproduce \cref{sec:scale} on the newer stack: Qwen3-8B 16.9 / 30.3 / 29.1 / 32.8\% against 16.7 / 30.3 / 27.4 / 33.2\% (\G/\E/\T/\ET), Qwen3-32B 31.3 / 35.3 / 34.4 / 35.1 against 32.0 / 35.7 / 34.0 / 35.9, Gemma~4 E4B 18.8 / 25.4 / 29.9 / 34.4 against 16.7 / 28.5 / 26.8 / 29.7. \cref{tab:scalecurve} and \figref{fig:scale} give the results; per-bucket accuracy and tool use for every size are in \cref{tab:scale_bucket,tab:scale_tools}.

\begin{table}[t]
\centering\small
\caption{\textbf{\run{ScaleCurve}: \facc\ (\%) on all 500 tasks by model size and tool set.} Left block: strict \facc, the per-task mean over the $T{=}0.7$ seeds (four for the Qwen3 sizes, two for Qwen3-14B-FP8 and Gemma), on the 485 tasks with function-level gold, with the paired arm$-$\G\ difference and its 95\% bootstrap interval; every Qwen difference has Holm-adjusted $p=0.002$. Right block: the same under canonical rescoring (not available for Gemma E2B/E4B, whose notebook kept only the last model's rescoring). Gemma~4 12B is one greedy run on the 307 tasks it completed; the run is degenerate (\cref{sec:app_scale}) and is listed for completeness only.}
\label{tab:scalecurve}
\setlength{\tabcolsep}{3pt}
\resizebox{\linewidth}{!}{%
\begin{tabular}{lr@{\hskip 1em}cccc@{\hskip 0.8em}lll@{\hskip 1.2em}cc@{\hskip 0.8em}l}
\toprule
 & & \multicolumn{7}{c}{strict \facc} & \multicolumn{3}{c}{canonical ids}\\
\cmidrule(lr){3-9}\cmidrule(lr){10-12}
Model & seeds & \G & \E & \T & \ET & \E$-$\G & \T$-$\G & \ET$-$\G & \G & \ET & \ET$-$\G\\
\midrule
Qwen3-0.6B & 4 & 0.2 & 9.4 & 7.0 & 8.0 & $+9.2$ \ci{7.6}{11.0} & $+6.8$ \ci{5.2}{8.4} & $+7.8$ \ci{6.4}{9.3} & 1.9 & 9.5 & $+7.6$ \ci{6.1}{9.1}\\
Qwen3-1.7B & 4 & 0.9 & 22.4 & 12.3 & 19.0 & $+21.5$ \ci{18.5}{24.6} & $+11.4$ \ci{9.4}{13.6} & $+18.1$ \ci{15.7}{20.7} & 11.2 & 24.5 & $+13.3$ \ci{10.7}{16.0}\\
Qwen3-4B (hybrid) & 4 & 10.1 & 27.7 & 27.1 & 29.5 & $+17.6$ \ci{14.6}{20.7} & $+17.0$ \ci{14.0}{20.0} & $+19.4$ \ci{16.4}{22.6} & 17.7 & 33.5 & $+15.7$ \ci{12.9}{18.6}\\
Qwen3-8B & 4 & 17.1 & 30.4 & 30.2 & 32.1 & $+13.3$ \ci{10.4}{16.3} & $+13.0$ \ci{10.2}{16.1} & $+15.0$ \ci{12.2}{17.9} & 25.7 & 34.3 & $+8.7$ \ci{5.9}{11.4}\\
Qwen3-14B & 4 & 32.1 & 44.0 & 41.4 & \textbf{46.0} & $+11.9$ \ci{8.7}{15.2} & $+9.3$ \ci{6.5}{12.2} & $+13.9$ \ci{11.0}{17.0} & 42.1 & \textbf{48.9} & $+6.9$ \ci{4.4}{9.4}\\
Qwen3-32B (FP8) & 4 & 32.1 & 35.8 & 37.1 & 36.5 & $+3.8$ \ci{1.7}{5.9} & $+5.1$ \ci{2.9}{7.3} & $+4.4$ \ci{2.3}{6.7} & 42.2 & 44.4 & $+2.2$ \ci{0.2}{4.3}\\
\midrule
Qwen3-14B-FP8 & 2 & 31.4 & 44.1 & 40.8 & 45.5 & $+12.7$ \ci{9.1}{16.3} & $+9.4$ \ci{6.4}{12.4} & $+14.0$ \ci{10.7}{17.4} & 40.8 & 48.0 & $+7.2$ \ci{4.4}{10.0}\\
\midrule
Gemma 4 E2B & 2 & 4.3 & 26.0 & 24.8 & 33.0 & $+21.6$ \ci{18.4}{25.1} & $+20.5$ \ci{17.1}{24.0} & $+28.7$ \ci{24.9}{32.6} & --- & --- & ---\\
Gemma 4 E4B & 2 & 18.0 & 28.2 & 29.1 & 32.4 & $+10.2$ \ci{7.5}{12.9} & $+11.0$ \ci{8.2}{13.8} & $+14.3$ \ci{11.4}{17.2} & --- & --- & ---\\
Gemma 4 12B (greedy, 307 tasks) & 0 & 17.3 & 18.6 & 19.3 & 18.3 & --- & --- & --- & 28.5 & 25.1 & ---\\
\bottomrule
\end{tabular}}
\end{table}

\textbf{The tools' gain over \grep\ rises from 0.6B to 1.7--4B and then falls monotonically to 32B.} Under strict scoring the best structured arm adds 9 points at 0.6B (where \G\ is at 0.2\% and the best arm at 9.4\%), 21 at 1.7B, 19 at 4B, 15 at 8B, 14 at 14B and 5 at 32B (\cref{tab:scalecurve}); all 18 Qwen differences are positive with Holm-adjusted $p=0.002$. A GEE model over the Qwen seeds, $y\sim\mathrm{arm}\times\log_2(\text{params})$ clustered by task, puts the trend at an odds ratio of 0.77 \ci{0.75}{0.79} per doubling for \E, 0.82 \ci{0.80}{0.84} for \T\ and 0.80 \ci{0.78}{0.82} for \ET\ (all $p<10^{-40}$): each doubling of parameters removes about a fifth of the structured tools' odds advantage, which is the estimate of \cref{sec:scale} (0.76--0.78) on six sizes and five decodings instead of three sizes and one. Adding hop as a control moves the estimates to 0.75 / 0.81 / 0.78.

\textbf{From 8B up, about half of the strict gain is the form of the submission.} Canonical rescoring raises the \G\ arm far more than the structured arms: \G\ goes from 17.1 to 25.7\% at 8B, 32.1 to 42.1\% at 14B and 32.1 to 42.2\% at 32B, while \ET\ moves by 2--3 points. The structured arms' gain over \G\ therefore shrinks from $+15.0$ to $+8.7$ (8B), $+13.9$ to $+6.9$ (14B) and $+4.4$ to $+2.2$ (32B), \figref{fig:scale}c. A larger model with \grep\ alone increasingly finds the target but names it in a form the strict matcher rejects (a line number, a \texttt{::} path, a bare short name), which is the form effect of \cref{sec:b5} measured across scale: the tools hand the model exact ids, and from 8B up that is about half of what they add. Below 8B the two scores agree more closely ($+19.4$ versus $+15.7$ at 4B; $+7.8$ versus $+7.6$ at 0.6B) because the small models' \G\ arm does not find the target in any form.

\textbf{The fall from 14B to 32B is a fall in tool uptake, not a rise in \grep\ competence.} \cref{sec:scale} read the small 32B gain as the larger model obtaining from \grep\ what the tools give the small ones. The curve refines that reading. Qwen3-14B's \G\ arm is exactly as accurate as Qwen3-32B's (32.1 versus 32.1\% strict, 42.1 versus 42.2\% canonical), yet the 14B model still gains 9--14 points (4--7 canonical) from the tools and the 32B model 4--5 (2). What differs is whether the tools are called: the 14B model calls \findentity\ in 96\% of its \T\ episodes and \semsearch\ in 95\% of its \E\ episodes; the 32B model in 45\% and 44\% (\figref{fig:scale}d). The FP8 control rules out precision as the cause: Qwen3-14B-FP8 matches Qwen3-14B arm by arm on the shared seeds (differences of $-1.6$ \ci{-3.5}{+0.4}, $+0.1$, $-0.7$ and $+0.4$ points for \G/\E/\T/\ET, all $p\ge0.14$) and gains as much from the tools (9--14 points). The uptake curve also shows which tool survives scale: \graphnb\ is called in 88\% of \T\ episodes at 8B but 13\% at 14B and 35\% at 32B, while \findentity\ is called in 80--98\% of \T\ episodes from 0.6B to 14B and only drops at 32B. Name lookup is the part of the graph tool that a larger model keeps using; neighbourhood expansion is the part it abandons first, and at 32B 1 of 1{,}498 \graphnb\ calls asks for two hops (8{,}424 of 9{,}653 at 8B). In absolute terms Qwen3-14B is the strongest model in the sweep (46.0\% with \ET; 48.9\% canonical) and the non-thinking Qwen3-32B is not better than it at this task in this harness (36.5\%; 44.4\% canonical), so the 32B point of \cref{sec:scale} should be read as a point about that checkpoint's agentic behaviour rather than as the asymptote of the curve.

\textbf{The depth gradient is present at every size and no arm flattens it.} With the arm as a control, each hop multiplies the odds of correct localization by 0.67 \ci{0.58}{0.78} at 0.6B, 0.58 at 1.7B, 0.57 at 4B, 0.53 at 8B, 0.50 at 14B and 0.60 \ci{0.52}{0.71} at 32B, every $p<10^{-6}$ (\cref{tab:scale_bucket}). In the joint model the $\hop\times\log_2(\text{params})$ interaction is 0.97 \ci{0.95}{1.00} ($p=0.033$): the gradient steepens very slightly with size, as the larger models' gains concentrate at hop 0 and hop 1. The $\hop\times\T$ interaction is never significantly below 1 at any size (0.83 at 8B, $p=0.06$; 0.86 at 14B, $p=0.03$, \ie\ \emph{steeper} than \G; 0.92 at 32B, $p=0.11$); the $\hop\times\E$ interaction is above 1 at 4B (1.71, $p=2{\times}10^{-4}$) and 32B (1.16, $p=0.012$). The pattern of \cref{sec:b1}---the graph arm's slope is no flatter than \grep's and the embedding arm's is the flattest---holds over the whole curve.

\textbf{What the small models do with the tools.} At 0.6B every arm is near the floor; the tools raise discovery (a gold function on screen) from 5\% of \G\ episodes to 42\% of \ET\ episodes but conversion stays at 16\%, and the \T\ arm issues 7.0 \findentity\ calls per episode and submits in only 52\%. At 1.7B and at the hybrid 4B the \G\ arm is the \grep\ loop of \cref{sec:b5}: 9.5 and 14.2 \grep\ calls per episode, 13\% and 33\% of episodes at the step cap, 76\% and 57\% submitting; the embedding arm ends the loop (6.1 steps and 95\% submission at 4B) and that behavioural change, not only the retrieval, is part of its gain. The Gemma family has the same shape as Qwen: E2B gains 21--29 points from a \G\ arm at 4.3\%, E4B 10--14 from one at 18.0\%. Gemma~4 12B's greedy run is degenerate (51\% of episodes at the step cap, 45\% submitting, 15.5 \readfile\ calls and 18k tokens per episode) and every arm scores 17--19\% on the 307 tasks it completed; we draw nothing from it (\cref{sec:app_scale}).

\textbf{The checkpoint matters as much as the size.} On the same 500 tasks with the same harness, \texttt{Qwen3-4B-Instruct-2507} reaches 26.8\% with \grep\ alone and gains nothing from embedding search (\cref{tab:full500}); the hybrid \texttt{Qwen3-4B} reaches 9.1\% and gains $+17.6$ points from it. The Instruct checkpoint's post-training makes it a competent \grep\ agent (7.6 steps, 98\% submission) that a semantic index does not help; the hybrid checkpoint is a looping \grep\ agent that any structured tool rescues. Which tools help is a property of how a checkpoint behaves as an agent, not of its parameter count, which is the strongest form of the per-model validation point of \cref{sec:discussion}. Finally, the planned non-inferiority test---can a 4B model with all tools match the 32B model with \grep\ alone?---comes out just short: hybrid Qwen3-4B \ET\ is $-2.5$ points \ci{-5.6}{+0.8} from Qwen3-32B \G\ on the shared seeds, and its one-sided 95\% lower bound ($-5.2$) misses the 5-point margin; \T\ alone is $-5.0$ \ci{-8.1}{-1.8}. Tools let a 4B model get within a few points of a 32B model's \grep\ arm, not past it.

%% file: sections/probes.tex
\section{Part 3: Does the model represent the repository graph?}
\label{sec:probes}

The tool ablation says what an agent can extract from structure that is handed to it. We now ask whether the structure is already in the model's activations, which is what the Thousand-Graph Hypothesis \citep{thousandgraph} asserts, and the kind of question tool-call-graph probing \citep{toolcallprobe} answers for a different graph. We probe for three relations between definitions $a$ and $b$ that both appear in the context: \textsc{call} ($a$ calls $b$ directly), \textsc{call2} ($a$ reaches $b$ in exactly two call hops and $b$'s short name does not occur in $a$'s source, so that the lexical cue is removed) and \textsc{inherit} ($a$'s class inherits from $b$). Negatives are pairs with no call or inheritance path of length $\le2$ in either direction; pairs where one definition contains the other are excluded.

\subsection{Representations}
\label{sec:repr}
A context is tokenised and run through the model in a single causal forward pass. Each definition $v$ in the context is represented by the residual-stream vector at its \emph{last} token, $\mathbf{h}^{(\ell)}_v\in\mathbb{R}^{d}$ at layer $\ell$: in a causal model the name token precedes the body, so only the end of the span has read what the definition calls. For a pair $(a,b)$ the probe sees
\begin{equation}
\mathbf{x}_{ab}=\big[\mathbf{h}_a;\ \mathbf{h}_b;\ \mathbf{h}_a\odot\mathbf{h}_b\big]\in\mathbb{R}^{3d},
\label{eq:pairfeat}
\end{equation}
and the one-sided controls see $\mathbf{h}_a$ or $\mathbf{h}_b$ alone; the interaction-only probe sees $\mathbf{h}_a\odot\mathbf{h}_b$. Features are standardised with training-set statistics, and a linear logistic probe $\sigma(\mathbf{w}^{\top}\mathbf{x}+c)$ is trained by Adam (learning rate $10^{-2}$, weight decay $10^{-2}$, 300 full-batch steps) on the class-weighted cross-entropy with positive weight $n_-/n_+$. We probe every fourth layer plus the last (Qwen3-4B and Qwen3-8B both have 36 layers; layer 0 is the token embedding). Folds are repository-held-out: train on the eleven non-Django repositories and test on Django, and the reverse.

\subsection{A global pair probe, and why it is not enough (\run{GlobalProbe})}
\label{sec:a1}
\textbf{Contexts.} For each of the 468 bucketed tasks we take the first gold-patched file at the base commit (one context per distinct file; 468 contexts, mean 4{,}668 tokens, truncated at 6{,}000), with a mean of 34.2 definitions each. Labels come from the whole-repository graph of \cref{sec:graph}, not just the file. This yields 13{,}255 \textsc{call}, 6{,}405 \textsc{call2} and 998 \textsc{inherit} positives; negatives are sampled at three per positive.

\textbf{Result.} The pair probe decodes \textsc{call} edges at AUC 0.785 (Django fold) and 0.747 (others) at layer 24 for Qwen3-4B, and 0.768/0.726 for Qwen3-8B, against 0.49--0.53 for a probe trained on permuted labels and 0.52--0.54 at layer 0 (\cref{tab:a1}, appendix). \textsc{call2} reaches 0.70/0.67 and \textsc{inherit} 0.77/0.80 at later layers. Taken alone these numbers would support a strong reading. The controls do not. A probe that sees only $\mathbf{h}_a$ reaches 0.71/0.66 on \textsc{call} and \emph{beats} the pair probe on \textsc{call2} (0.70/0.69 vs 0.70/0.67) and on \textsc{inherit} (0.79/0.80 vs 0.77/0.81); a probe that sees only $\mathbf{h}_b$ reaches 0.61/0.61 on \textsc{call} and 0.75/0.76 on \textsc{inherit}. A non-neural baseline, the in-file in-degree of $b$, scores 0.85/0.85 on \textsc{call} and 0.74/0.76 on \textsc{call2}, and whether $b$'s name occurs in $a$'s text scores 0.97 on \textsc{call}. Positive and negative \emph{pairs} differ in properties of a \emph{single} definition: callers are long, callees are popular, base classes are base-class-like. A global AUC therefore says little about whether the model represents the \emph{relation}.

\subsection{Within-anchor ranking with degree stratification (\run{AnchorProbe})}
\label{sec:a1c}
To remove node-level shortcuts we score each pair only against other pairs sharing the same anchor. For every anchor $a$ that has at least one positive partner, the candidate set is its positives and up to 15 of its negatives from the same context. Let $\mathcal{A}$ be the set of groups, each with positives $P_g$ and negatives $N_g$, and let $s(\cdot)$ be a score. The metric is the pooled within-group AUC,
\begin{equation}
\AUC_{\mathrm{grp}}(s)=\frac{\sum_{g\in\mathcal{A}}\sum_{p\in P_g}\sum_{n\in N_g}\big(\mathbf{1}[s(p)>s(n)]+\tfrac12\mathbf{1}[s(p)=s(n)]\big)}{\sum_{g\in\mathcal{A}}|P_g|\,|N_g|},
\label{eq:groupauc}
\end{equation}
the probability that a random positive outranks a random negative \emph{of the same anchor}. Any feature of $a$ alone is constant within a group and cannot help. Under the \emph{strict} grouping, the group is the anchor crossed with the in-degree bin of $b$ (bins $0$, $1$, $2$, $3$--$4$, $5{+}$), so that the popularity of the candidate is held fixed as well; the in-degree baseline then scores 0.50--0.53 by construction. Controls are: the $\mathbf{h}_b$-only probe (any remaining node-level signal about $b$), the interaction-only probe, a pair probe trained on labels permuted \emph{within} each training group, the in-degree baseline, token distance, same-class membership, and the lexical baseline $\mathbf{1}[\text{name}(b)\in\text{text}(a)]$. Per-pair test scores are saved at layers 24 and 36, and 95\% intervals are obtained by bootstrapping anchors (2{,}000 resamples; the numerator and denominator of \cref{eq:groupauc} are resampled jointly).

\begin{table}[t]
\centering\small
\caption{\textbf{\run{AnchorProbe}: within-anchor AUC at layer 24}, strict grouping (anchor $\times$ in-degree bin of $b$), with 95\% anchor-bootstrap intervals for Qwen3-4B. A relational representation shows as pair/interaction AUC above $b$-only, the permuted control and the in-degree baseline. The lexical baseline is $\mathbf{1}[\text{name}(b)\in\text{text}(a)]$; it is 0.5 for \textsc{call2} by construction. Qwen3-8B point estimates from the same protocol are given in the last two columns (pair / $b$-only).}
\label{tab:a1c}
\setlength{\tabcolsep}{4pt}
\resizebox{\linewidth}{!}{%
\begin{tabular}{llcccccc@{\hskip 1em}cc}
\toprule
 & & \multicolumn{6}{c}{Qwen3-4B} & \multicolumn{2}{c}{Qwen3-8B}\\
\cmidrule(lr){3-8}\cmidrule(lr){9-10}
Task & Test fold & pair & interaction & $b$-only & permuted & in-degree & lexical & pair & $b$-only\\
\midrule
\textsc{call} & Django & 0.740 \ci{.724}{.756} & 0.731 \ci{.715}{.746} & 0.510 \ci{.493}{.527} & 0.502 \ci{.484}{.520} & 0.534 & 0.993 & 0.733 & 0.533\\
 & others & 0.683 \ci{.665}{.700} & 0.669 \ci{.651}{.686} & 0.530 \ci{.511}{.550} & 0.497 \ci{.479}{.516} & 0.513 & 0.987 & 0.661 & 0.523\\
\textsc{call2} & Django & 0.569 \ci{.548}{.590} & 0.532 \ci{.512}{.552} & 0.520 \ci{.497}{.542} & 0.476 \ci{.458}{.493} & 0.501 & 0.500 & 0.606 & 0.608\\
 & others & 0.573 \ci{.552}{.593} & 0.555 \ci{.534}{.577} & 0.559 \ci{.536}{.581} & 0.509 \ci{.489}{.529} & 0.500 & 0.500 & 0.552 & 0.562\\
\textsc{inherit} & Django & 0.619 \ci{.593}{.647} & 0.569 \ci{.544}{.597} & 0.688 \ci{.663}{.715} & 0.462 \ci{.436}{.487} & 0.500 & 0.927 & 0.655 & 0.667\\
 & others & 0.629 \ci{.588}{.672} & 0.501 \ci{.461}{.546} & 0.674 \ci{.633}{.714} & 0.471 \ci{.432}{.511} & 0.500 & 0.893 & 0.655 & 0.725\\
\bottomrule
\end{tabular}}

\end{table}

\begin{figure}[t]
\centering
\includegraphics[width=\linewidth]{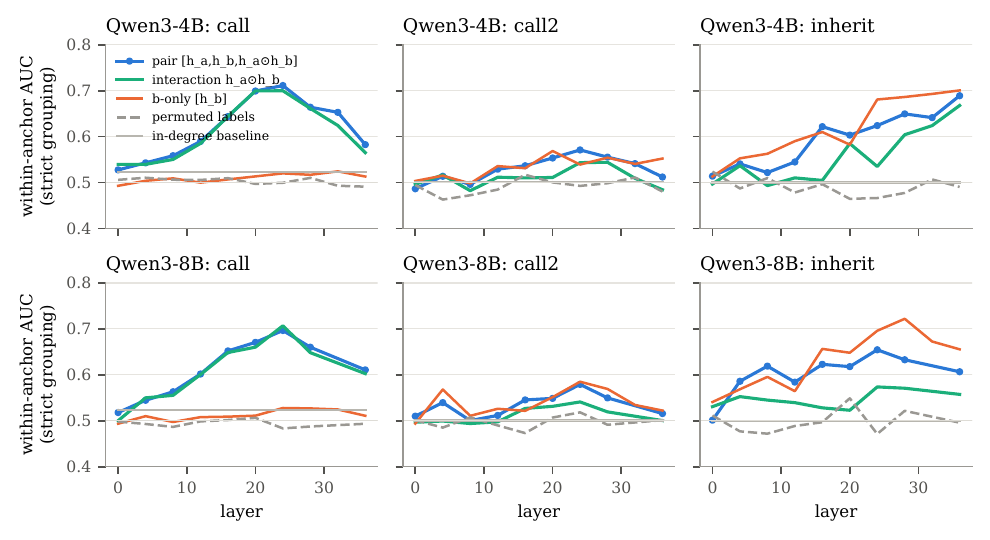}
\caption{\textbf{\run{AnchorProbe} across layers} (strict grouping, mean of the two folds). Direct \textsc{call} edges are decodable from the middle layers on, peaking at layer 24, with the interaction term carrying almost all of the signal and the $b$-only probe at chance; \textsc{call2} is barely above $b$-only; for \textsc{inherit} the $b$-only probe is at or above the pair probe. The degenerate Qwen3-8B layer-32 fit (AUC exactly 0.5 for every pair probe, a numerical failure of the standardised logistic fit at that layer) is omitted.}
\label{fig:probelayers}
\end{figure}

\cref{tab:a1c} and \figref{fig:probelayers} give the result, which is sharply task-dependent.

\textbf{Direct call edges are linearly represented.} Within anchor and in-degree bin, the pair probe ranks a function's true callees above its non-callees with AUC 0.740 $\ci{0.724}{0.756}$ (Django held out) and 0.683 $\ci{0.665}{0.700}$ (others held out) for Qwen3-4B, and 0.733/0.661 for Qwen3-8B, while the $b$-only probe is at 0.51--0.53, the permuted control at 0.50 and in-degree at 0.51--0.53. The interaction-only probe $\mathbf{h}_a\odot\mathbf{h}_b$ is almost as good (0.731/0.669), so the signal is a bilinear compatibility between caller and callee rather than anything additive. It emerges in the middle of the network (0.54--0.56 at layers 4--8, 0.62--0.74 at layers 16--24) and decays towards the output (0.57--0.59 at layer 36). The relation survives the controls that defeated the global probe, and it transfers across repositories.

\textbf{But it is weaker than a one-line lexical baseline.} Whether $b$'s name appears in $a$'s source text scores 0.993/0.987 on the same groups. The model's representation of ``$a$ calls $b$'' is a degraded copy of information that is literally in the text of $a$; nothing in this experiment distinguishes a learned relational code from a smeared-out lexical match.

\textbf{Two-hop edges with the lexical cue removed are barely decodable.} For \textsc{call2}, where $b$'s name does not occur in $a$'s text, the pair probe scores 0.569/0.573 and the $b$-only probe 0.520/0.559; the pair--$b$-only margin is 0.05 on the Django fold and 0.01 on the other, with overlapping intervals on the latter. For Qwen3-8B the pair probe does not beat $b$-only at all (0.606 vs 0.608, 0.552 vs 0.562). What two-hop signal exists is at most a few points above what the callee alone predicts.

\textbf{Inheritance is not decodable beyond one-sided information.} The pair probe (0.619/0.629) is \emph{below} the $b$-only probe (0.688/0.674) in both folds and both models, and the interaction-only probe is at 0.50 on the others fold: base classes are recognisable as base classes, but which class inherits from which is not read out by a linear probe, even though the base name is in the \texttt{class} line and the lexical baseline scores 0.89--0.93.

\subsection{The same probes inside agent conversations (\run{ConvProbe})}
\label{sec:a2}
A static file is not what an agent sees. We repeat the within-anchor protocol on activations taken inside real conversations: the 159 greedy Qwen3-4B episodes of arms \G\ (50) and \T\ (109) from \run{Abl-4B} in which at least four probe-able entities appear. Each conversation is rendered exactly as the model saw it (chat template plus the arm's tool schemas; mean 5{,}178 tokens, maximum 31{,}578, cap 32{,}768) and run through the model once; a definition is probe-able if its name is unique in the snapshot and its \texttt{def}/\texttt{class} line appears in the conversation, and it is represented at the last token of its \emph{last} occurrence. Pairs are labelled from the repository graph as before (\textsc{call}: 1{,}078 positives, 9{,}510 negatives; \textsc{call2}: 801 and 4{,}846) and additionally tagged as \emph{stated} if some \graphnb\ output line contains both ids, \ie\ a tool printed that exact edge. Arm \T\ contributes most pairs (17.9 probe-able entities per conversation against 7.8 for \G), since its tools print definition lines.

\begin{table}[t]
\centering\small
\caption{\textbf{\run{ConvProbe}: within-anchor AUC inside agent conversations} (Qwen3-4B, layer 24, strict grouping), with anchor-bootstrap intervals, for subsets of the test pairs. \emph{unstated} pairs are those for which no graph tool printed the edge. Strict groups cannot be formed for the \emph{stated} subset (almost all stated pairs are positives), so that row reports the plain within-anchor grouping.}
\label{tab:a2}
\setlength{\tabcolsep}{4pt}
\resizebox{\linewidth}{!}{%
\begin{tabular}{llrrcccc}
\toprule
Task & Subset (fold) & $n$ & pos & pair & $b$-only & permuted & in-degree\\
\midrule
\textsc{call} & all (Django) & 4860 & 500 & 0.605 \ci{.513}{.690} & 0.511 \ci{.423}{.594} & 0.458 \ci{.365}{.554} & 0.521\\
 & all (others) & 5728 & 578 & 0.655 \ci{.590}{.717} & 0.520 \ci{.447}{.595} & 0.516 \ci{.455}{.578} & 0.580\\
 & unstated (Django) & 4565 & 246 & 0.590 \ci{.474}{.697} & 0.479 \ci{.370}{.580} & 0.486 \ci{.374}{.596} & 0.503\\
 & unstated (others) & 5399 & 294 & 0.599 \ci{.515}{.686} & 0.483 \ci{.390}{.584} & 0.529 \ci{.446}{.613} & 0.541\\
 & arm \G\ (Django) & 238 & 33 & 0.778 \ci{.500}{1.00} & 0.556 \ci{.125}{.889} & 0.333 \ci{.000}{.667} & 0.500\\
 & arm \G\ (others) & 396 & 64 & 0.706 \ci{.467}{.900} & 0.588 \ci{.312}{.840} & 0.588 \ci{.294}{.833} & 0.529\\
 & stated (Django)$^\ast$ & 295 & 254 & 0.850 \ci{.646}{1.00} & 0.900 \ci{.739}{1.00} & 0.500 \ci{.314}{.660} & 0.875\\
 & stated (others)$^\ast$ & 329 & 284 & 0.744 \ci{.576}{.893} & 0.846 \ci{.703}{.959} & 0.410 \ci{.244}{.595} & 1.000\\
\textsc{call2} & all (Django) & 2218 & 298 & 0.747 \ci{.691}{.802} & 0.781 \ci{.728}{.831} & 0.448 \ci{.391}{.509} & 0.500\\
 & all (others) & 3429 & 503 & 0.645 \ci{.583}{.701} & 0.647 \ci{.587}{.703} & 0.461 \ci{.403}{.519} & 0.501\\
\bottomrule
\multicolumn{8}{l}{\footnotesize $^\ast$plain within-anchor grouping.}
\end{tabular}}

\end{table}

\cref{tab:a2} shows that the static picture carries over, somewhat attenuated. Within anchor and in-degree bin, direct \textsc{call} edges are decodable at 0.605 $\ci{0.513}{0.690}$ and 0.655 $\ci{0.590}{0.717}$ against 0.51--0.52 for $b$-only and 0.46--0.52 for the permuted control. The intervals are wide because a conversation exposes far fewer definitions than a file (fewer and smaller groups), but they exclude the controls on the ``others'' fold and nearly do on Django. The restriction to \emph{unstated} pairs---edges no tool ever printed---gives 0.590/0.599 against 0.48 for $b$-only: the model's ranking of callees does not depend on a \graphnb\ line having asserted the edge, which is consistent with the static result that the representation tracks the text of the caller, since a \readfile\ or \grep\ hit on the caller's body is what usually brings both definitions into context. On the \emph{stated} pairs the $b$-only probe scores as high as the pair probe (0.90/0.85), because a stated callee is simply a callee that appeared in a \graphnb\ line, which is a property of $b$ in that conversation. For \textsc{call2}, the pair probe is again no better than $b$-only (0.747 vs 0.781; 0.645 vs 0.647).

In-conversation, as in static code, the model carries a linearly readable but shallow and largely lexical representation of \emph{direct} dependencies, and no linearly readable representation of two-hop dependencies beyond what the callee alone predicts. The 32k-token conversation that a \T\ episode produces does not leave behind a decodable graph; it leaves behind the same direct-call signal, weaker.

%% file: sections/discussion.tex
\section{Discussion}
\label{sec:discussion}

\paragraph{One picture from three parts.}
The three parts of the study were designed independently but say the same thing. The hardest third of SWE-bench Verified---hop 2 and hop 3+, 154 tasks---is hard for every arm of every model at every size from 0.6B to 32B, and the tools that are supposed to address exactly this difficulty do not. The graph tools' contribution is concentrated where the issue names the target (hop 0, where \findentity\ is a direct lookup) or its immediate neighbour (hop 1, where one \graphnb\ expansion suffices); past that, agents rarely chain expansions (12--14\% of episodes) and the arm's depth slope is at least as steep as \grep's. The embedding tool's contribution is flatter in depth because semantic similarity does not care about graph distance, but it is small in absolute terms (13--28\% of first reveals) and depends on retrieval quality. Inside the model, the only relation that is linearly readable beyond node-level confounds is the direct call edge, and even that is a weaker signal than a substring test on the caller's source. Two-hop structure is essentially absent from the activations, is rarely assembled from the tools, and is where the tasks are failed. Small coding agents operate with a one-hop picture of the repository, in weights and in use. The scale curve adds a fourth strand: what the tools add is largest for the models that cannot \grep\ competently (1.7--4B), shrinks by about a fifth of the odds advantage per doubling, is half submission form from 8B up, and disappears at 32B because the model stops calling the tools, with neighbourhood expansion abandoned before name lookup. The structure that is extracted is the one-hop part, and it is extracted less as the model grows.

\paragraph{On the contrast with LocAgent.}
LocAgent \citep{locagent} reports that graph-guided agents degrade least with depth. Our graph arm degrades at least as steeply as \grep\ alone. The difference is not the graph (ours follows LocAgent's construction) and is unlikely to be the measurement (our hop label is deterministic and robust to resolution). The plausible differences are model scale and agent design: LocAgent's results use larger models and a tool that returns compressed subgraphs with relation-aware summaries, whereas our \T\ arm hands a 4--8B model raw edge lines truncated at 50 and leaves the chaining to it. Scale alone does not close the gap in our harness: at none of six sizes from 0.6B to 32B is the graph arm's depth slope flatter than the \grep\ arm's (\cref{sec:scalecurve}), and the larger models simply call the graph tools less. \run{ForceTool} rules out ``the agent does not call the tool'' as the explanation, and \run{Budget} rules out budget. What is left is the model's ability to \emph{use} a neighbourhood listing to decide what to expand next, which is a reasoning step that, by the probe results, has no linear support in the model's representation of the code. We think this is the right reading, but it is an inference, and a direct test---presenting multi-hop paths explicitly, or training the agent to chain expansions---is future work.

\paragraph{What the probes do and do not show.}
Our protocol is deliberately conservative. The within-anchor, degree-binned AUC removes every node-level shortcut we could name, and the lexical baseline calibrates what ``decodable'' means against a trivial alternative. A linear probe is also a lower bound: a nonlinear readout, an attention-pattern analysis, or a causal intervention could find two-hop structure a linear probe cannot. The claim we make is correspondingly narrow: \emph{linearly}, at the \emph{definition's last token}, the repository's call graph beyond one hop is not represented in these models, and the one-hop part that is represented is not more than lexical. The agreement between the static probe and the in-conversation probe, including on edges the tools never printed, adds confidence that this is a property of the model and not of the context format. We also want to be explicit about the negative finding behind the design: the global pair-probe AUC of 0.75--0.80 that we obtained first, and that a less careful study would have reported, is matched by one-sided probes and by a degree baseline. We recommend the within-anchor protocol, with a stratification on the confound that matters for the relation (here, in-degree), for any relational probe.

\paragraph{Implications for tool design.}
Three practical suggestions follow. (i) For small models, exact-name resolution is the single most valuable thing a structured tool provides, both for discovery at hop 0 and for the \emph{form} of the submission; it is cheap and should be present even when a full graph is not. (ii) Since agents expand once and stop, a tool that returns the \emph{two-hop} closure in a compact, ranked form, or that proposes the next expansion, is more likely to help at hop 2+ than a tool that returns raw one-hop edges. (iii) Embedding search and graph tools reveal the gold through different routes and combine additively for the 8B model and only marginally for Gemma; for Qwen3-4B-Instruct-2507 the embedding tool is dead weight and \ET\ is slightly worse than \T, while for the hybrid Qwen3-4B checkpoint of the same size it is worth $+17.6$ points. Tool sets should be validated per checkpoint rather than inherited from a larger model or from another checkpoint of the same size. (iv) Beyond about 8B, a strict scorer credits the tools for the form of the submission as much as for discovery; a localization stage that canonicalises names (line numbers, \texttt{::} paths, short names) recovers about half of the tools' measured gain without any tool.

\paragraph{Implications for benchmarks.}
Hop depth is a cheap, released, validated axis along which to report results. Aggregate \facc\ on SWE-bench Verified mixes 37\% of tasks where the answer is named in the issue with 15\% where it is three or more edges away; a method that improves the former and not the latter will look good on the aggregate and has not addressed structure. We suggest reporting the hop-3+ bucket separately, and we note that the strict-mention variant, though it discards too many tasks to be the default, is a useful conservative check on any claim about deep tasks.

\paragraph{Limitations.}
The primary models are small (4--8B) and the mechanism experiments (\run{Traces}, \run{Budget}, \run{ForceTool}, \run{Repair}, the probes) are on the 4--8B models only. The scale curve covers 0.6B--32B with five decodings per size, but its largest model is the non-thinking Qwen3-32B served from an FP8 checkpoint, which is not stronger than Qwen3-14B at this task; the curve should not be read as an asymptote, and whether a model that calls the tools \emph{and} is strong at \grep\ would show a different curve is open. The Gemma family contributes two usable sizes (E2B, E4B); the 12B run was partial and degenerate. Canonical rescoring is available for the Qwen sizes only, and the scale curve's 4B model is a different checkpoint from the primary 4B runs, which we treat as a finding about checkpoints rather than a confound, but which means the curve has no point for the Instruct checkpoint at 4B. The evaluation subset is 160 tasks, which limits the resolution of arm-versus-arm comparisons among the structured tools to about $\pm5$ points even with three runs; Django supplies half of the subset. The hop label is optimistic (name-based resolution, static only) and sensitive to mention extraction; depth is a shortest path, not an agent's path. The label and the \T\ arm's tools are built from the same graph, so the hop-0/hop-1 advantage of \T\ is in part an advantage \emph{on the label's own terms}: a target the label calls ``one edge away'' is one that \graphnb\ can print, and a resolution rule that drops an ambiguous call drops it for both. The \E\ and \G\ arms do not share this coupling, which makes the \E\ results the cleaner test of the depth axis. The localization-only formulation removes the interaction between exploration and editing that a full agent has, and the \run{Repair} model is itself weak, so \run{Repair} bounds rather than measures the value of localization. The probes use one representation (last-token residual stream) and one family of readouts (linear); the in-conversation probe has wide intervals and is on the 4B model only. The embedding arm uses one strong and one weak embedder, not a sweep. Finally, the study was conducted on one benchmark in one language; Python's dynamic dispatch is invisible to our graph, which both understates true depth and overstates what a static graph tool could ever return.

%% file: sections/conclusion.tex
\section{Conclusion}
\label{sec:conclusion}
We labelled SWE-bench Verified with a deterministic code-graph hop depth, showed that each hop roughly halves the odds of correct function-level localization for four models from 4B to 32B after controls and at every one of six sizes from 0.6B to 32B, and released the labels. We ablated the tool set of a localization agent and found that structured tools roughly double \grep-only accuracy for Qwen3-8B and Gemma~4 E4B, that only graph tools help Qwen3-4B, that over a scale curve from 0.6B to 32B the tools' gain peaks at 17--21 points at 1.7--4B and falls to 4--5 at 32B as the model stops calling them (with about half of the gain from 8B up being the form of the submission), that the same size gains or does not gain from a tool depending on the checkpoint, and that the graph tools' benefit is exact-name lookup at hop 0 and one expansion at hop 1---they do not flatten the depth gradient, and forcing their use, enlarging the budget or sampling does not change that. A fixed repair model turns those localizations into 6--13 resolved tasks of 160 against 26 with oracle locations, almost all of them on correctly localized tasks. Linear probes, once stratified to remove node-level confounds, find a direct-call representation in the middle layers that transfers across repositories and survives inside agent conversations, but that is weaker than a substring test and does not extend to two hops or to inheritance. The multi-hop structure that the deepest third of the benchmark requires is neither in these models' weights nor, in practice, extracted from the tools we gave them. Closing that gap---by tools that return and rank multi-hop closures, by training agents to chain expansions, or by models that represent more than one hop---is the open problem this study isolates.

%% file: sections/appendix.tex
\section{Hop depth: additional detail}
\label{sec:app_depth}

\begin{table}[H]
\centering\small
\caption{\textbf{Hop-depth buckets by repository} (default label, all 500 SWE-bench Verified tasks).}
\label{tab:byrepo}
\begin{tabular}{lrrrrrrr}
\toprule
Repository & hop 0 & hop 1 & hop 2 & hop 3+ & no mention & unreachable & total\\
\midrule
django/django & 87 & 64 & 29 & 38 & 13 & 0 & 231\\
sympy/sympy & 28 & 25 & 14 & 5 & 3 & 0 & 75\\
sphinx-doc/sphinx & 2 & 5 & 11 & 17 & 9 & 0 & 44\\
matplotlib/matplotlib & 14 & 8 & 8 & 4 & 0 & 0 & 34\\
scikit-learn/scikit-learn & 20 & 9 & 3 & 0 & 0 & 0 & 32\\
astropy/astropy & 10 & 6 & 5 & 1 & 0 & 0 & 22\\
pydata/xarray & 13 & 5 & 3 & 1 & 0 & 0 & 22\\
pytest-dev/pytest & 7 & 4 & 2 & 5 & 1 & 0 & 19\\
pylint-dev/pylint & 2 & 1 & 0 & 2 & 4 & 1 & 10\\
psf/requests & 2 & 1 & 2 & 2 & 1 & 0 & 8\\
mwaskom/seaborn & 1 & 0 & 1 & 0 & 0 & 0 & 2\\
pallets/flask & 0 & 0 & 0 & 1 & 0 & 0 & 1\\
\midrule
all & 186 & 128 & 78 & 76 & 31 & 1 & 500\\
\bottomrule
\end{tabular}
\end{table}

\paragraph{Worked example.} The labeller's self-test illustrates the definition. In a four-file package where \texttt{c.py:Child(Base)} defines \texttt{go()} which calls \texttt{self.run()}, \texttt{a.py:Base.run} calls \texttt{helper()} imported from \texttt{b.py}, and \texttt{b.py:helper} calls \texttt{b.py:deep}, a patch to \texttt{deep} with the issue ``Calling \texttt{Child.go()} returns the wrong value'' has mention $\{\texttt{Child}\}$ (\texttt{go} is too short to count) and depth 4: \texttt{Child} $\xrightarrow{\textsc{contain}}$ \texttt{Child.go} $\xrightarrow{\textsc{invoke}}$ \texttt{Base.run} $\xrightarrow{\textsc{invoke}}$ \texttt{helper} $\xrightarrow{\textsc{invoke}}$ \texttt{deep}, bucket hop 3+; under containment edges only it is unreachable. The issue ``\texttt{deep} is broken'' has depth 0.

\paragraph{Mention patterns.} The six regular expressions are, in order:
\begin{lstlisting}
`([^`\n]{2,80})`                                  back-ticked span
\b([A-Za-z_][\w\.]*)\s*\(                         identifier before a parenthesis
File "[^"]*?([\w/]+\.py)", line \d+, in (\w+)     traceback frame
\b([A-Za-z_]\w*(?:\.[A-Za-z_]\w*)+)\b             dotted name
\b([a-z]+_[a-z_0-9]+)\b                           snake_case
\b([A-Z][a-z0-9]+[A-Z]\w*)\b                      CamelCase
\end{lstlisting}
Every captured group is further split into identifier-like pieces (\verb|[A-Za-z_][\w/]*\.py| or \verb|[A-Za-z_]\w*|). The strict variant keeps only the first and third patterns.

\section{GEE models: arm effects, interactions and per-arm slopes}
\label{sec:app_gee}

\begin{table}[H]
\centering\small
\caption{\textbf{Additional GEE estimates} (default label, with controls). Top: arm effects from \cref{eq:gee}. Middle: the interaction model $y\sim\hop\times\mathrm{arm}+\mathrm{repo}+n^{\mathrm{gold}}+\log(1+L)$; the interaction OR multiplies the per-hop OR of \G\ to give that arm's per-hop OR. Bottom: per-hop OR from separate fits on each arm's episodes. ``Qwen3-8B (3)'' pools the greedy run and the two sampled runs.}
\label{tab:gee_extra}
\setlength{\tabcolsep}{4pt}
\resizebox{\linewidth}{!}{%
\begin{tabular}{llccccc}
\toprule
 & Term & Qwen3-4B (3 runs) & Qwen3-8B (greedy) & Qwen3-8B (3) & Gemma 4 E4B (3) & Qwen3-32B (500)\\
\midrule
Arm effect & \E\ vs \G & 0.93 \ci{0.72}{1.20} & 2.83 \ci{1.74}{4.61} & 2.61 \ci{1.78}{3.81} & 2.48 \ci{1.78}{3.45} & 1.21 \ci{1.01}{1.44}\\
(OR) & \T\ vs \G & 1.68 \ci{1.20}{2.37} & 2.40 \ci{1.46}{3.94} & 2.53 \ci{1.72}{3.71} & 2.06 \ci{1.48}{2.88} & 1.13 \ci{0.94}{1.36}\\
 & \ET\ vs \G & 1.42 \ci{1.00}{2.02} & 3.43 \ci{2.00}{5.89} & 2.96 \ci{1.98}{4.45} & 2.80 \ci{1.98}{3.95} & 1.24 \ci{1.03}{1.49}\\
 & per gold function & 0.49 \ci{0.23}{1.07} & 0.72 \ci{0.38}{1.38} & 0.71 \ci{0.34}{1.47} & 0.44 \ci{0.15}{1.25} & 0.67 \ci{0.39}{1.15}\\
 & per $\log$ patch line & 0.86 \ci{0.51}{1.47} & 0.58 \ci{0.31}{1.06} & 0.59 \ci{0.31}{1.10} & 1.01 \ci{0.53}{1.94} & 0.67 \ci{0.47}{0.94}\\
\midrule
Interaction & hop (arm \G) & 0.46 \ci{0.31}{0.69} & 0.46 \ci{0.26}{0.82} & 0.46 \ci{0.27}{0.76} & 0.45 \ci{0.30}{0.67} & 0.55 \ci{0.44}{0.70}\\
(OR) & hop$\times$\E & 1.24 \ci{1.00}{1.54}, $p{=}.048$ & 1.05 \ci{0.62}{1.77} & 1.22 \ci{0.79}{1.87} & 1.30 \ci{0.95}{1.77}, $p{=}.098$ & 1.13 \ci{0.95}{1.36}\\
 & hop$\times$\T & 0.84 \ci{0.59}{1.20} & 0.93 \ci{0.54}{1.58} & 0.89 \ci{0.59}{1.36} & 1.01 \ci{0.74}{1.38} & 0.94 \ci{0.78}{1.15}\\
 & hop$\times$\ET & 1.08 \ci{0.79}{1.49} & 1.07 \ci{0.61}{1.88} & 1.01 \ci{0.63}{1.63} & 1.12 \ci{0.77}{1.61} & 0.91 \ci{0.75}{1.10}\\
\midrule
Per-arm & \G & 0.47 \ci{0.31}{0.70} & 0.43 \ci{0.24}{0.74} & 0.41 \ci{0.23}{0.71} & 0.39 \ci{0.27}{0.55} & 0.56 \ci{0.44}{0.71}\\
hop OR & \E & 0.60 \ci{0.43}{0.85} & 0.45 \ci{0.29}{0.69} & 0.55 \ci{0.38}{0.78} & 0.57 \ci{0.42}{0.78} & 0.62 \ci{0.50}{0.77}\\
 & \T & 0.34 \ci{0.22}{0.53} & 0.36 \ci{0.22}{0.60} & 0.35 \ci{0.22}{0.55} & 0.45 \ci{0.31}{0.65} & 0.52 \ci{0.41}{0.67}\\
 & \ET & 0.52 \ci{0.36}{0.73} & 0.49 \ci{0.32}{0.75} & 0.43 \ci{0.29}{0.63} & 0.48 \ci{0.33}{0.68} & 0.49 \ci{0.39}{0.62}\\
\bottomrule
\end{tabular}}

\end{table}

\section{Agent results: per-run and per-bucket detail}
\label{sec:app_runs}

\begin{table}[H]
\centering\small
\caption{\textbf{\facc\ (\%) by hop bucket for every run.} 40 tasks per bucket (38--40 with function-level gold). \run{Budget} is on 20 tasks per bucket.}
\label{tab:bybucket_runs}
\setlength{\tabcolsep}{3.5pt}
\resizebox{\linewidth}{!}{%
\begin{tabular}{llcccc@{\hskip 1em}cccc@{\hskip 1em}cccc@{\hskip 1em}cccc}
\toprule
 & & \multicolumn{4}{c}{\G} & \multicolumn{4}{c}{\E} & \multicolumn{4}{c}{\T} & \multicolumn{4}{c}{\ET}\\
\cmidrule(lr){3-6}\cmidrule(lr){7-10}\cmidrule(lr){11-14}\cmidrule(lr){15-18}
Model & Run & 0 & 1 & 2 & 3+ & 0 & 1 & 2 & 3+ & 0 & 1 & 2 & 3+ & 0 & 1 & 2 & 3+\\
\midrule
Qwen3-4B & \run{Abl-4B} & 45.0 & 26.3 & 15.0 & 10.5 & 40.0 & 18.4 & 15.0 & 18.4 & 55.0 & 31.6 & 22.5 & 13.2 & 47.5 & 28.9 & 27.5 & 13.2\\
 & \run{Seeds} s1 & 37.5 & 31.6 & 15.0 & 10.5 & 30.0 & 15.8 & 22.5 & 10.5 & 62.5 & 23.7 & 25.0 & 10.5 & 50.0 & 23.7 & 27.5 & 15.8\\
 & \run{Seeds} s2 & 37.5 & 21.1 & 12.5 & 7.9 & 40.0 & 18.4 & 20.0 & 7.9 & 57.5 & 28.9 & 25.0 & 10.5 & 45.0 & 23.7 & 20.0 & 10.5\\
 & \run{ForceTool} & --- & --- & --- & --- & 60.0 & 18.4 & 22.5 & 7.9 & 52.5 & 23.7 & 22.5 & 18.4 & --- & --- & --- & ---\\
\midrule
Qwen3-8B & \run{Abl-8B} & 25.0 & 15.8 & 5.0 & 5.3 & 47.5 & 23.7 & 20.0 & 13.2 & 47.5 & 18.4 & 20.0 & 7.9 & 55.0 & 18.4 & 27.5 & 15.8\\
 & \run{Seeds} s1 & 25.0 & 15.8 & 2.5 & 5.3 & 37.5 & 21.1 & 17.5 & 10.5 & 52.5 & 21.1 & 12.5 & 7.9 & 50.0 & 21.1 & 20.0 & 13.2\\
 & \run{Seeds} s2 & 32.5 & 13.2 & 5.0 & 5.3 & 45.0 & 21.1 & 17.5 & 18.4 & 47.5 & 26.3 & 20.0 & 5.3 & 52.5 & 21.1 & 12.5 & 7.9\\
 & \run{ForceTool} & --- & --- & --- & --- & 52.5 & 13.2 & 22.5 & 15.8 & 47.5 & 21.1 & 15.0 & 2.6 & --- & --- & --- & ---\\
 & \run{WeakEmb} & --- & --- & --- & --- & 47.5 & 15.8 & 12.5 & 7.9 & --- & --- & --- & --- & --- & --- & --- & ---\\
 & \run{Budget} & 15.0 & 21.1 & 10.0 & 5.3 & --- & --- & --- & --- & --- & --- & --- & --- & 55.0 & 31.6 & 30.0 & 15.8\\
\midrule
Gemma 4 E4B & \run{Abl-Gemma} & 20.0 & 10.5 & 7.5 & 0.0 & 25.0 & 31.6 & 15.0 & 15.8 & 40.0 & 23.7 & 12.5 & 10.5 & 47.5 & 21.1 & 12.5 & 13.2\\
 & \run{Seeds} s1 & 27.5 & 13.2 & 10.0 & 7.9 & 45.0 & 34.2 & 17.5 & 15.8 & 42.5 & 23.7 & 10.0 & 7.9 & 52.5 & 34.2 & 12.5 & 18.4\\
 & \run{Seeds} s2 & 37.5 & 18.4 & 7.5 & 0.0 & 42.5 & 34.2 & 15.0 & 10.5 & 50.0 & 23.7 & 12.5 & 10.5 & 57.5 & 18.4 & 10.0 & 26.3\\
\midrule
Qwen3-32B & \run{Abl-32B}, subset & 40.0 & 36.8 & 15.0 & 15.8 & 50.0 & 34.2 & 20.0 & 23.7 & 45.0 & 31.6 & 20.0 & 15.8 & 45.0 & 36.8 & 12.5 & 21.1\\
\bottomrule
\end{tabular}}

\end{table}

\begin{table}[H]
\centering\small
\caption{\textbf{Pre-fix run (\run{Abl-4B-prefix}, Qwen3-4B).} The first full run, before the forced-tool-call rule of \cref{sec:harness}; 10 episodes errored and 148 tasks have all four arms. Reported for completeness only; the absolute level and the arm differences are both confounded by episodes that spent their budget repeating a tool-less message.}
\label{tab:prefix}
\begin{tabular}{lcccccc}
\toprule
Arm & \facc\ & func.\ recall@10 & file Acc@5 & steps & submit & $n$\\
\midrule
\G & 17.5 & 20.8 & 38.0 & 11.8 & 0.797 & 158\\
\E & 20.0 & 23.4 & 43.4 & 13.7 & 0.736 & 159\\
\T & 19.6 & 21.7 & 38.2 & 15.4 & 0.637 & 157\\
\ET & 22.4 & 26.6 & 40.4 & 14.8 & 0.692 & 156\\
\bottomrule
\end{tabular}
\end{table}

\begin{table}[H]
\centering\small
\caption{\textbf{First tool to reveal a gold function, by bucket} (fraction of episodes; ``none'' = gold never revealed). Qwen panels pool three runs. Abbreviations: fe = \findentity, gn = \graphnb, ss = \semsearch, rf = \readfile.}
\label{tab:firsttool}
\setlength{\tabcolsep}{3pt}
\resizebox{\linewidth}{!}{%
\begin{tabular}{llcccccc@{\hskip 1em}cccccc@{\hskip 1em}cccc}
\toprule
 & & \multicolumn{6}{c}{arm \T} & \multicolumn{6}{c}{arm \ET} & \multicolumn{4}{c}{arm \E}\\
\cmidrule(lr){3-8}\cmidrule(lr){9-14}\cmidrule(lr){15-18}
Model & bucket & grep & rf & fe & gn & none & & grep & rf & fe & gn & ss & none & grep & rf & ss & none\\
\midrule
Qwen3-4B & hop 0 & .28 & .01 & .53 & .07 & .12 & & .32 & .03 & .53 & .03 & .01 & .09 & .78 & .07 & .03 & .13\\
 & hop 1 & .29 & .07 & .04 & .40 & .20 & & .22 & .08 & .06 & .40 & .01 & .23 & .30 & .20 & .07 & .43\\
 & hop 2 & .35 & .03 & .04 & .17 & .41 & & .29 & .07 & .03 & .14 & .02 & .46 & .30 & .07 & .13 & .49\\
 & hop 3+ & .17 & .04 & .00 & .16 & .64 & & .17 & .03 & .03 & .15 & .04 & .60 & .17 & .09 & .13 & .60\\
\midrule
Qwen3-8B & hop 0 & .32 & .03 & .46 & .12 & .08 & & .42 & .03 & .25 & .03 & .17 & .10 & .59 & .03 & .20 & .17\\
 & hop 1 & .25 & .02 & .02 & .49 & .22 & & .26 & .10 & .03 & .27 & .08 & .25 & .40 & .16 & .15 & .30\\
 & hop 2 & .18 & .09 & .05 & .17 & .50 & & .15 & .09 & .04 & .17 & .11 & .44 & .16 & .09 & .28 & .47\\
 & hop 3+ & .12 & .01 & .01 & .16 & .70 & & .07 & .04 & .00 & .17 & .13 & .60 & .13 & .05 & .13 & .68\\
\midrule
Gemma 4 E4B & hop 0 & .42 & .12 & .28 & .08 & .10 & & .35 & .16 & .07 & .04 & .35 & .03 & .38 & .16 & .42 & .04\\
 & hop 1 & .15 & .07 & .03 & .43 & .33 & & .19 & .12 & .00 & .21 & .22 & .25 & .21 & .15 & .29 & .35\\
 & hop 2 & .33 & .12 & .01 & .06 & .48 & & .12 & .12 & .00 & .08 & .23 & .45 & .15 & .10 & .24 & .51\\
 & hop 3+ & .23 & .10 & .00 & .05 & .62 & & .14 & .04 & .00 & .06 & .23 & .53 & .14 & .06 & .25 & .54\\
\midrule
Qwen3-32B (500) & hop 0 & .63 & .06 & .18 & .03 & .10 & & .49 & .10 & .29 & .01 & .01 & .10 & .67 & .09 & .09 & .16\\
 & hop 1 & .37 & .27 & .03 & .13 & .20 & & .31 & .28 & .06 & .08 & .02 & .25 & .33 & .26 & .09 & .32\\
 & hop 2 & .32 & .20 & .00 & .05 & .43 & & .29 & .14 & .01 & .01 & .00 & .54 & .18 & .17 & .07 & .58\\
 & hop 3+ & .30 & .11 & .03 & .01 & .55 & & .19 & .15 & .00 & .00 & .06 & .60 & .11 & .14 & .07 & .69\\
\bottomrule
\end{tabular}}

\end{table}

\paragraph{Transitions.}\sloppy The most frequent tool-to-tool transitions per episode in the pooled traces are: Qwen3-4B \G: \grep$\to$\grep\ 2.03, \grep$\to$\readfile\ 1.24, \readfile$\to$\grep\ 0.63; Qwen3-4B \T: \grep$\to$\grep\ 0.96, start$\to$\findentity\ 0.82, \findentity$\to$\graphnb\ 0.75, \graphnb$\to$\grep\ 0.55; Qwen3-8B \G: \grep$\to$\grep\ 8.53, \readfile$\to$\readfile\ 2.88, \listdir$\to$\listdir\ 2.32; Qwen3-8B \T: \grep$\to$\grep\ 3.41, \readfile$\to$\readfile\ 1.94, \graphnb$\to$\graphnb\ 1.44, \findentity$\to$\graphnb\ 0.99; Qwen3-8B \E: \grep$\to$\grep\ 5.68, \grep$\to$\semsearch\ 0.55, \semsearch$\to$\grep\ 0.55. The median step at which the gold is first revealed is 1 for every Qwen arm (0--2 for Gemma~4 E4B), so when an agent finds the gold it usually finds it immediately; the failures are episodes that never do, or that see it and submit something else.

\section{Probes: additional tables}
\label{sec:app_probes}

\begin{table}[H]
\centering\small
\caption{\textbf{\run{GlobalProbe}: global pair-probe AUC at layer 24} with one-sided probes and non-neural baselines on the test fold. ``best'' gives the best layer for the pair probe and its AUC. Positives/negatives per test fold: \textsc{call} 7{,}918/36{,}291 (Django) and 5{,}337/25{,}523 (others); \textsc{call2} 3{,}526/33{,}985 and 2{,}879/23{,}993; \textsc{inherit} 690/36{,}291 and 308/25{,}523. For \textsc{call2} both the lexical cue and the name baseline are removed by construction.}
\label{tab:a1}
\setlength{\tabcolsep}{3.5pt}
\resizebox{\linewidth}{!}{%
\begin{tabular}{llcccccccccc}
\toprule
Model & Task / fold & pair & permuted & $a$-only & $b$-only & name & $-$distance & same class & in-degree($b$) & out-degree($a$) & best\\
\midrule
Qwen3-4B & \textsc{call} / Django & .785 & .520 & .709 & .613 & .968 & .609 & .561 & .847 & .893 & L24 .785\\
 & \textsc{call} / others & .747 & .487 & .656 & .608 & .970 & .604 & .538 & .852 & .860 & L24 .747\\
 & \textsc{call2} / Django & .700 & .514 & .695 & .591 & .500 & .507 & .542 & .744 & .832 & L24 .700\\
 & \textsc{call2} / others & .672 & .509 & .687 & .614 & .500 & .477 & .512 & .763 & .797 & L24 .672\\
 & \textsc{inherit} / Django & .711 & .499 & .743 & .708 & .968 & .677 & .392 & .292 & .306 & L36 .774\\
 & \textsc{inherit} / others & .708 & .477 & .772 & .688 & .970 & .687 & .368 & .301 & .283 & L28 .805\\
\midrule
Qwen3-8B & \textsc{call} / Django & .768 & .497 & .690 & .603 & .968 & .609 & .561 & .847 & .893 & L24 .768\\
 & \textsc{call} / others & .726 & .525 & .667 & .602 & .970 & .604 & .538 & .852 & .860 & L24 .726\\
 & \textsc{call2} / Django & .654 & .541 & .670 & .619 & .500 & .507 & .542 & .744 & .832 & L20 .673\\
 & \textsc{call2} / others & .661 & .497 & .697 & .616 & .500 & .477 & .512 & .763 & .797 & L24 .661\\
 & \textsc{inherit} / Django & .699 & .501 & .753 & .727 & .968 & .677 & .392 & .292 & .306 & L24 .699\\
 & \textsc{inherit} / others & .703 & .427 & .785 & .724 & .970 & .687 & .368 & .301 & .283 & L28 .761\\
\bottomrule
\end{tabular}}

\end{table}

\begin{table}[H]
\centering\small
\caption{\textbf{\run{AnchorProbe} under the plain within-anchor grouping} (anchor only, no in-degree stratification), Qwen3-4B, layer 24, with anchor-bootstrap intervals. Compared with \cref{tab:a1c}, the $b$-only probe and the in-degree baseline recover a large part of their score, which is exactly the node-level signal the strict grouping removes. Anchors with a positive: 2{,}937 / 2{,}650 (\textsc{call}), 1{,}478 / 1{,}342 (\textsc{call2}), 631 / 299 (\textsc{inherit}).}
\label{tab:a1c_plain}
\setlength{\tabcolsep}{4pt}
\resizebox{\linewidth}{!}{%
\begin{tabular}{llcccccc}
\toprule
Task & Test fold & pair & interaction & $b$-only & permuted & in-degree & lexical\\
\midrule
\textsc{call} & Django & 0.742 \ci{.732}{.753} & 0.727 \ci{.716}{.737} & 0.622 \ci{.614}{.631} & 0.523 \ci{.514}{.532} & 0.892 \ci{.885}{.899} & 0.991\\
 & others & 0.705 \ci{.693}{.717} & 0.672 \ci{.660}{.685} & 0.617 \ci{.606}{.627} & 0.498 \ci{.488}{.508} & 0.889 \ci{.881}{.897} & 0.985\\
\textsc{call2} & Django & 0.595 \ci{.583}{.608} & 0.567 \ci{.554}{.579} & 0.595 \ci{.582}{.608} & 0.500 \ci{.487}{.513} & 0.790 \ci{.776}{.805} & 0.500\\
 & others & 0.612 \ci{.600}{.624} & 0.554 \ci{.540}{.568} & 0.626 \ci{.612}{.641} & 0.519 \ci{.506}{.532} & 0.789 \ci{.778}{.801} & 0.500\\
\textsc{inherit} & Django & 0.628 \ci{.605}{.651} & 0.572 \ci{.548}{.596} & 0.681 \ci{.659}{.704} & 0.484 \ci{.462}{.506} & 0.301 \ci{.291}{.313} & 0.891\\
 & others & 0.607 \ci{.569}{.644} & 0.490 \ci{.452}{.530} & 0.660 \ci{.624}{.697} & 0.455 \ci{.419}{.490} & 0.325 \ci{.305}{.344} & 0.862\\
\bottomrule
\end{tabular}}

\end{table}

\paragraph{Layer 36.} At the final layer the \textsc{call} signal has decayed: strict within-anchor pair AUC 0.591 $\ci{0.570}{0.610}$ (Django) and 0.574 $\ci{0.555}{0.594}$ (others) for Qwen3-4B, with $b$-only at 0.51 and the control at 0.49--0.50. The relation is a mid-network property, consistent with the layer profile of \figref{fig:probelayers}.

\paragraph{\run{ConvProbe} across layers.} For \textsc{call} under the strict grouping (all pairs), the in-conversation pair probe is at 0.54/0.47 at layer 0, 0.56/0.52 at layer 8, 0.54/0.59 at layer 16, 0.61/0.66 at layer 24, 0.66/0.61 at layer 32 and 0.60/0.62 at layer 36 (Django/others), with the $b$-only probe between 0.44 and 0.56 throughout. The plain (non-strict) within-anchor pair AUC peaks at 0.81/0.83 at layer 24, against 0.80/0.79 for $b$-only and 0.97/0.95 for in-degree, which repeats the static lesson that the unstratified metric is dominated by node-level signal.

\section{\run{ScaleCurve}: additional tables}
\label{sec:app_scale}

\cref{tab:scale_bucket} gives strict \facc\ by hop bucket for every model and arm of \run{ScaleCurve} with the per-size depth gradient, and \cref{tab:scale_tools} the episode outcomes, costs and tool-call counts behind \figref{fig:scale}d.

\paragraph{Gemma 4 12B.} The serving session ended during the greedy run after 307 tasks (1{,}228 episodes; 120/82/45/45/15 tasks at hop 0/1/2/3+/no-mention, all four arms for each). Unlike every other model in the sweep the 12B agent loops on \readfile\ (15.5--16.9 calls per episode, 18--20k tokens), hits the 30-step cap in 51--57\% of episodes and submits in 39--45\%, so every arm scores 17--19\% (strict) with the gold on screen in 88\% of episodes and a conversion rate of 20--22\%. Whether this is a property of the checkpoint, of the gemma4 tool-call parser in vLLM 0.32.0, or of the session is not something the partial run can settle; we report it and draw no conclusion from it.

\paragraph{Replication of the primary runs.} Three models were run both in the primary experiments (\cref{sec:scale}; vLLM 0.19.1) and in \run{ScaleCurve} (vLLM 0.32.0, path-fixed harness). Greedy strict \facc\ on all 500 tasks, primary versus \run{ScaleCurve}: Qwen3-8B 16.7/30.3/27.4/33.2 versus 16.9/30.3/29.1/32.8; Qwen3-32B-FP8 32.0/35.7/34.0/35.9 versus 31.3/35.3/34.4/35.1; Gemma~4 E4B 16.7/28.5/26.8/29.7 versus 18.8/25.4/29.9/34.4 (\G/\E/\T/\ET). The Qwen runs agree within 2 points per arm; the Gemma E4B structured arms differ by 3--5 points between stacks, which is within the between-seed spread of that model (\cref{tab:b4}) but should be kept in mind when the two Gemma E4B numbers are compared.

\paragraph{Missing traces.} In six runs the per-episode results file is complete but the full message-trace file stopped early (Qwen3-4B seed 2: 1{,}240 of 2{,}000 traces; Qwen3-8B seed 1: 1{,}732; Qwen3-14B seeds 1 and 2: 1{,}168 and 300; Qwen3-32B seed 2: 1{,}864; Qwen3-14B-FP8 seed 1: 1{,}832; Gemma E4B seed 1: 228); the Drive mount stopped accepting writes to the trace file while the results file kept growing. Every metric in this section comes from the results file (which records each step's tool, arguments, revealed nodes and token counts), so none is affected.

\begin{table}[H]
\centering\footnotesize\setlength{\tabcolsep}{3.5pt}
\caption{\textbf{\run{ScaleCurve}: strict \facc\ (\%) by hop bucket, model and arm} ($T{=}0.7$ seeds pooled; Gemma~4 12B greedy). The last column is the per-hop odds ratio from a GEE fit $y\sim\mathrm{arm}+\hop$ clustered by task on that model's seed episodes (Qwen3 sizes only), with its 95\% interval.}
\label{tab:scale_bucket}
\begin{tabular}{ll@{\hskip 1em}cccc@{\hskip 1.2em}l}
\toprule
Model & arm & hop 0 & hop 1 & hop 2 & hop 3+ & OR per hop\\
\midrule
Qwen3-0.6B & \G & 0.4 & 0.2 & 0.0 & 0.0 & 0.67 \ci{0.57}{0.78}\\
 & \E & 9.9 & 13.0 & 7.6 & 4.5 & \\
 & \T & 11.2 & 9.3 & 1.6 & 0.3 & \\
 & \ET & 10.8 & 8.5 & 4.6 & 2.7 & \\
Qwen3-1.7B & \G & 1.8 & 0.6 & 0.3 & 0.0 & 0.58 \ci{0.49}{0.69}\\
 & \E & 30.3 & 22.6 & 14.5 & 11.0 & \\
 & \T & 21.9 & 13.0 & 1.6 & 1.7 & \\
 & \ET & 28.6 & 19.3 & 7.9 & 7.2 & \\
Qwen3-4B (hybrid) & \G & 18.8 & 9.3 & 1.6 & 1.4 & 0.57 \ci{0.47}{0.68}\\
 & \E & 37.8 & 29.5 & 16.4 & 14.0 & \\
 & \T & 45.5 & 28.9 & 8.6 & 3.1 & \\
 & \ET & 41.4 & 35.8 & 10.9 & 14.7 & \\
Qwen3-8B & \G & 25.1 & 24.2 & 5.6 & 2.7 & 0.53 \ci{0.43}{0.64}\\
 & \E & 44.1 & 30.9 & 15.8 & 15.8 & \\
 & \T & 49.2 & 30.1 & 12.8 & 7.5 & \\
 & \ET & 47.4 & 35.8 & 13.8 & 12.7 & \\
Qwen3-14B & \G & 47.7 & 38.4 & 12.8 & 8.6 & 0.50 \ci{0.42}{0.60}\\
 & \E & 64.7 & 43.9 & 21.7 & 21.9 & \\
 & \T & 61.9 & 47.6 & 19.7 & 8.2 & \\
 & \ET & 63.9 & 50.0 & 22.0 & 24.0 & \\
Qwen3-32B (FP8) & \G & 42.8 & 38.4 & 15.8 & 14.4 & 0.60 \ci{0.52}{0.70}\\
 & \E & 45.4 & 40.7 & 20.4 & 24.0 & \\
 & \T & 51.9 & 42.1 & 17.8 & 16.8 & \\
 & \ET & 48.0 & 43.7 & 15.8 & 21.9 & \\
\midrule
Qwen3-14B-FP8 & \G & 47.0 & 40.2 & 9.9 & 6.2 & \\
 & \E & 63.2 & 45.9 & 21.7 & 25.3 & \\
 & \T & 61.4 & 46.7 & 16.4 & 12.3 & \\
 & \ET & 64.3 & 47.6 & 24.3 & 23.3 & \\
\midrule
Gemma 4 E2B & \G & 7.3 & 4.1 & 3.3 & 0.0 & \\
 & \E & 38.6 & 24.0 & 14.5 & 13.0 & \\
 & \T & 41.1 & 27.2 & 6.6 & 3.4 & \\
 & \ET & 51.1 & 33.7 & 11.8 & 11.6 & \\
Gemma 4 E4B & \G & 24.9 & 24.0 & 5.9 & 5.5 & \\
 & \E & 37.8 & 30.5 & 19.1 & 15.1 & \\
 & \T & 41.4 & 36.2 & 11.8 & 8.9 & \\
 & \ET & 43.0 & 39.0 & 17.8 & 15.8 & \\
Gemma 4 12B (greedy) & \G & 21.0 & 17.7 & 11.6 & 9.5 & \\
 & \E & 21.0 & 21.5 & 11.6 & 14.3 & \\
 & \T & 21.8 & 25.3 & 7.0 & 9.5 & \\
 & \ET & 22.7 & 19.0 & 9.3 & 11.9 & \\
\bottomrule
\end{tabular}
\end{table}

\begin{table}[H]
\centering\footnotesize\setlength{\tabcolsep}{3pt}
\caption{\textbf{\run{ScaleCurve}: episode outcomes and tool use per model and arm}, all decodings pooled. \emph{saw} is the fraction of episodes in which a gold function was revealed, \emph{conv.} is $\Pr(\text{correct}\mid\text{saw})$, \emph{cap} the fraction at the 30-step cap; calls per episode for \grep, \readfile, \semsearch\ (ss), \findentity\ (fe) and \graphnb\ (gn); ``use'' is the fraction of episodes with at least one call to the structured tool (ss for \E; fe/gn for \T\ and \ET); ``2-hop'' is the fraction of \graphnb\ calls that ask for \texttt{hops=2}.}
\label{tab:scale_tools}
\resizebox{\linewidth}{!}{%
\begin{tabular}{llcccccc@{\hskip 1em}ccccc@{\hskip 1em}cc}
\toprule
Model & arm & acc & saw & conv. & submit & cap & steps & \grep & \readfile & ss & fe & gn & use & 2-hop\\
\midrule
Qwen3-0.6B & \G & 0.2 & 0.05 & 0.03 & 0.64 & 0.07 & 7.9 & 1.5 & 0.7 & --- & --- & --- & --- & ---\\
 & \E & 9.4 & 0.40 & 0.20 & 0.72 & 0.13 & 10.4 & 1.1 & 2.5 & 2.93 & --- & --- & .87 & ---\\
 & \T & 7.0 & 0.21 & 0.23 & 0.52 & 0.22 & 13.8 & 1.7 & 0.2 & --- & 7.00 & 0.11 & .98/.08 & .32\\
 & \ET & 8.0 & 0.42 & 0.16 & 0.71 & 0.16 & 12.3 & 1.1 & 1.3 & 4.44 & 1.48 & 0.44 & .68/.42 & .20\\
Qwen3-1.7B & \G & 0.9 & 0.28 & 0.02 & 0.76 & 0.13 & 13.8 & 9.5 & 0.5 & --- & --- & --- & --- & ---\\
 & \E & 22.4 & 0.46 & 0.45 & 0.97 & 0.02 & 7.2 & 3.8 & 0.2 & 1.01 & --- & --- & .84 & ---\\
 & \T & 12.3 & 0.45 & 0.25 & 0.73 & 0.23 & 17.6 & 6.7 & 0.6 & --- & 4.80 & 1.98 & .98/.48 & .12\\
 & \ET & 19.0 & 0.50 & 0.35 & 0.91 & 0.08 & 12.2 & 4.7 & 0.3 & 0.81 & 2.27 & 1.41 & .93/.47 & .11\\
Qwen3-4B (hybrid) & \G & 10.1 & 0.37 & 0.24 & 0.57 & 0.33 & 16.1 & 14.2 & 0.3 & --- & --- & --- & --- & ---\\
 & \E & 27.7 & 0.55 & 0.45 & 0.95 & 0.03 & 6.1 & 4.1 & 0.1 & 0.78 & --- & --- & .74 & ---\\
 & \T & 27.1 & 0.56 & 0.47 & 0.77 & 0.19 & 12.1 & 8.3 & 0.4 & --- & 1.13 & 0.96 & .80/.60 & .16\\
 & \ET & 29.5 & 0.63 & 0.46 & 0.91 & 0.08 & 8.7 & 5.2 & 0.2 & 0.54 & 0.72 & 0.86 & .58/.63 & .47\\
Qwen3-8B & \G & 17.1 & 0.57 & 0.30 & 0.67 & 0.33 & 15.3 & 8.6 & 3.0 & --- & --- & --- & --- & ---\\
 & \E & 30.4 & 0.68 & 0.44 & 0.84 & 0.16 & 11.6 & 5.1 & 3.1 & 1.47 & --- & --- & .88 & ---\\
 & \T & 30.2 & 0.68 & 0.44 & 0.79 & 0.20 & 13.5 & 4.7 & 2.9 & --- & 1.64 & 2.41 & .94/.88 & .85\\
 & \ET & 32.1 & 0.71 & 0.46 & 0.86 & 0.14 & 11.6 & 5.3 & 1.9 & 0.61 & 0.87 & 1.46 & .71/.79 & .91\\
Qwen3-14B & \G & 32.1 & 0.69 & 0.44 & 0.96 & 0.01 & 8.2 & 4.6 & 2.3 & --- & --- & --- & --- & ---\\
 & \E & 44.0 & 0.73 & 0.59 & 0.99 & 0.00 & 6.8 & 2.9 & 1.9 & 0.99 & --- & --- & .95 & ---\\
 & \T & 41.4 & 0.76 & 0.54 & 0.96 & 0.02 & 10.1 & 4.9 & 2.2 & --- & 1.48 & 0.20 & .96/.13 & .60\\
 & \ET & 46.0 & 0.76 & 0.59 & 0.99 & 0.00 & 8.5 & 3.3 & 1.9 & 0.68 & 1.22 & 0.29 & .90/.24 & .77\\
Qwen3-32B (FP8) & \G & 32.1 & 0.66 & 0.45 & 0.98 & 0.01 & 5.2 & 1.8 & 2.4 & --- & --- & --- & --- & ---\\
 & \E & 35.8 & 0.66 & 0.50 & 0.99 & 0.00 & 5.0 & 1.1 & 2.5 & 0.44 & --- & --- & .44 & ---\\
 & \T & 37.1 & 0.73 & 0.48 & 0.97 & 0.02 & 7.0 & 2.3 & 2.8 & --- & 0.53 & 0.42 & .45/.35 & .00\\
 & \ET & 36.5 & 0.70 & 0.47 & 0.98 & 0.01 & 6.0 & 1.5 & 2.6 & 0.12 & 0.53 & 0.18 & .47/.17 & .00\\
\midrule
Qwen3-14B-FP8 & \G & 31.4 & 0.69 & 0.45 & 0.95 & 0.01 & 8.5 & 4.6 & 2.5 & --- & --- & --- & --- & ---\\
 & \E & 44.1 & 0.73 & 0.58 & 0.99 & 0.00 & 6.9 & 2.9 & 2.0 & 0.99 & --- & --- & .95 & ---\\
 & \T & 40.8 & 0.76 & 0.53 & 0.96 & 0.02 & 10.4 & 5.1 & 2.4 & --- & 1.37 & 0.28 & .94/.18 & .52\\
 & \ET & 45.5 & 0.77 & 0.58 & 0.98 & 0.00 & 8.7 & 3.5 & 2.0 & 0.68 & 1.13 & 0.29 & .88/.24 & .81\\
\midrule
Gemma 4 E2B & \G & 4.3 & 0.48 & 0.07 & 0.96 & 0.04 & 8.7 & 3.0 & 2.6 & --- & --- & --- & --- & ---\\
 & \E & 26.0 & 0.63 & 0.38 & 0.96 & 0.03 & 6.1 & 0.9 & 2.7 & 1.08 & --- & --- & .99 & ---\\
 & \T & 24.8 & 0.62 & 0.38 & 0.96 & 0.04 & 9.2 & 1.5 & 2.4 & --- & 1.57 & 1.72 & .99/.88 & .08\\
 & \ET & 33.0 & 0.70 & 0.44 & 0.97 & 0.03 & 7.4 & 0.5 & 1.9 & 0.88 & 1.19 & 1.65 & .90/.90 & .05\\
Gemma 4 E4B & \G & 18.0 & 0.63 & 0.26 & 1.00 & 0.00 & 6.8 & 2.0 & 2.2 & --- & --- & --- & --- & ---\\
 & \E & 28.2 & 0.70 & 0.36 & 1.00 & 0.00 & 6.1 & 1.4 & 2.3 & 0.77 & --- & --- & .75 & ---\\
 & \T & 29.1 & 0.71 & 0.40 & 1.00 & 0.00 & 8.1 & 1.8 & 2.1 & --- & 0.85 & 0.92 & .69/.69 & .04\\
 & \ET & 32.4 & 0.74 & 0.43 & 1.00 & 0.00 & 7.0 & 1.2 & 2.2 & 0.72 & 0.22 & 1.00 & .20/.69 & .04\\
Gemma 4 12B (greedy) & \G & 17.3 & 0.88 & 0.20 & 0.45 & 0.51 & 21.7 & 3.6 & 15.5 & --- & --- & --- & --- & ---\\
 & \E & 18.6 & 0.89 & 0.21 & 0.43 & 0.54 & 22.1 & 3.4 & 16.2 & 0.03 & --- & --- & .02 & ---\\
 & \T & 19.3 & 0.89 & 0.21 & 0.40 & 0.57 & 23.0 & 4.6 & 15.7 & --- & 0.22 & 0.06 & .20/.06 & .00\\
 & \ET & 18.3 & 0.89 & 0.20 & 0.39 & 0.56 & 23.3 & 3.9 & 16.9 & 0.02 & 0.20 & 0.01 & .18/.01 & .00\\
\bottomrule
\end{tabular}}
\end{table}

\section{Prompts and tool schemas}
\label{sec:app_prompts}

\paragraph{Localization agent system prompt.}
\begin{lstlisting}
You are a code localization agent. Given a GitHub issue and a Python repository, find the functions, methods or classes that must be edited to resolve the issue. Explore with the tools, then call submit once with a ranked list of locations, most likely first, each as 'path/to/file.py:QualifiedName' (e.g. 'pkg/mod.py:Class.method'), or 'path/to/file.py' for module-level code. Do not write a patch.
\end{lstlisting}
The user turn is \texttt{Issue:\textbackslash n} followed by the first 6{,}000 characters of the problem statement. After a turn without a tool call the harness appends the user message \texttt{Use a tool, or call submit with your ranked locations.} and requires a tool call on the next turn.

\paragraph{Tool descriptions (as given in the JSON schemas).}
\begin{itemize}
\item \grep: ``Regex search over the repository's .py files. Returns up to 40 'path:line: text' matches.'' Parameters \texttt{pattern} (required), \texttt{path\_prefix}.
\item \readfile: ``Read lines of a file with line numbers (max 120 lines).'' Parameters \texttt{path} (required), \texttt{start}, \texttt{end}.
\item \listdir: ``List the files and subdirectories under a directory ('' = repository root).'' Parameter \texttt{path}.
\item \semsearch: ``Embedding search: the 10 functions/classes whose code is most similar in meaning to a natural-language or code query.'' Parameter \texttt{query} (required).
\item \findentity: ``Find functions/classes by name. Returns entity ids 'path:QualifiedName' with signatures.'' Parameter \texttt{name} (required).
\item \graphnb: ``Code-graph neighbours of an entity id or file path: what it contains, calls, imports, inherits, and what calls/imports/inherits/contains it. hops = 1 or 2.'' Parameters \texttt{entity} (required), \texttt{hops}.
\item \submit: ``Submit the final ranked list of locations and stop.'' Parameter \texttt{locations} (array of strings, required).
\end{itemize}

\paragraph{\run{Repair} system prompt.}
\begin{lstlisting}
You are fixing a bug in a Python repository. You are given the issue and the code of the functions that are most likely involved. Change only that code. Answer with one or more edit blocks and nothing else, each exactly:
<<<<<<< SEARCH path/to/file.py
<exact lines copied from the code shown>
=======
<replacement lines>
>>>>>>> REPLACE
The SEARCH lines must match the file exactly, including indentation.
\end{lstlisting}
The user turn contains the issue (first 6{,}000 characters) and, for each of up to five locations, a header \texttt{\#\#\# path (lines s--e): QualifiedName} followed by the first 150 lines of the definition in a code fence. Every arm's completion contained at least one well-formed edit block in 98--100\% of prompts (mean 1.1 blocks); a block is applied only if its SEARCH text occurs exactly once in the file, and the result is diffed against the base commit with \texttt{difflib}.

\section{Compute and software}
\label{sec:app_compute}
The hop-depth labelling, \run{Abl-4B}, \run{Abl-4B-prefix}, the Qwen3-4B halves of \run{Seeds} and \run{ForceTool}, the static probes (\run{GlobalProbe}, \run{AnchorProbe}) and the \run{Repair} harness scoring ran on a GCP \texttt{g2-standard-16} VM with one NVIDIA L4; \run{Abl-8B}, \run{Abl-Gemma}, the Qwen3-8B halves of \run{Seeds} and \run{ForceTool}, \run{Budget}, \run{WeakEmb}, the 8B probes, \run{Repair} generation and \run{ConvProbe} ran on Colab A100 40GB instances; \run{Abl-32B}, \run{Abl-8B-500} and the Gemma~4 E4B seeds of \run{Seeds} ran on Colab A100 80GB instances (the 32B model from \texttt{Qwen/Qwen3-32B-FP8}); the GEE fits, label variants, trace analysis, conversation rendering and the repair prompts and patches ran on CPU. Serving used vLLM (0.19.1) with \texttt{--max-model-len 32768} (40960 for \run{Budget}) and the hermes or gemma4 tool-call parser; probes used \texttt{transformers} in bfloat16 with SDPA attention; the SWE-bench harness was version 4.1.0 with prebuilt instance images, 6 workers and an 1{,}800\,s per-instance timeout. Embeddings were cached by text hash across base commits. A \run{Repair} scoring attempt that hit Docker image-name conflicts for part of the \G\ arm was re-run for the affected instances; no instance errored in the final reports. Checkpoints: \texttt{Qwen/Qwen3-4B-Instruct-2507}, \texttt{Qwen/Qwen3-8B}, \texttt{Qwen/Qwen3-32B-FP8}, \texttt{Qwen/Qwen3-Embedding-0.6B}, \texttt{minishlab/potion-base-8M}, \texttt{google/gemma-4-E4B-it}. \run{ScaleCurve} ran on Colab A100-SXM4-80GB instances with vLLM 0.32.0, transformers 5.19.0 and torch 2.13.0, one serving session per model with every arm and decoding of that model in the session, the same 30-step and 32k-token limits, and a build of the agent that differs from \texttt{scripts/loc\_agent.py} only in output ordering and resume handling (the harness and graph builder used are stored with the results, with md5s, and each session's \texttt{env\_*.json} records the pip freeze, GPU, model revisions and the exact \texttt{vllm serve} commands). Checkpoints: \texttt{Qwen/Qwen3-0.6B}, \texttt{Qwen3-1.7B}, \texttt{Qwen3-4B}, \texttt{Qwen3-8B}, \texttt{Qwen3-14B}, \texttt{Qwen3-14B-FP8}, \texttt{Qwen3-32B-FP8}, \texttt{google/gemma-4-E2B-it}, \texttt{gemma-4-E4B-it}, \texttt{gemma-4-12B-it}.